\documentclass[12pt]{article}

\usepackage[utf8]{inputenc}
\usepackage[english]{babel}

\usepackage{multicol}
\usepackage{comment}
\usepackage{natbib}
\usepackage{csquotes}
\usepackage{booktabs} 
\usepackage{multirow}
\usepackage{graphicx}
\usepackage{enumerate}
\usepackage{grffile} 
\usepackage{amsmath}
\usepackage{hyperref}
\usepackage{mathtools}
\usepackage{graphicx}
\usepackage{subcaption}
\usepackage{algorithm}
\usepackage{algpseudocode}
\usepackage{parskip}
\usepackage{geometry}
\usepackage{xcolor}

\def \R {\mathbb{R}}
\def \E {\mathbb{E}}

\usepackage{amssymb}
\usepackage{amsthm}
\theoremstyle{plain}

\newtheorem{theorem}{Theorem}
\numberwithin{theorem}{section} 

\numberwithin{proposition}{section} 
\newtheorem{assumption}{Assumption} 
\numberwithin{assumption}{section} 

\numberwithin{lemma}{section}
\newtheorem{corollary}{Corollary}
\numberwithin{corollary}{section}

\usepackage{titlesec}

\titleformat{\paragraph}
{\normalfont\normalsize\bfseries}{\theparagraph}{1em}{}
\titlespacing*{\paragraph}
{0pt}{3.25ex plus 1ex minus .2ex}{1.5ex plus .2ex}

\newif\ifarxiv
\arxivtrue 

\begin{document}

\title{\bf Single-Index Quantile Factor Model with Observed Characteristics}

\author{
  Ruofan Xu\thanks{
    This paper benefited greatly from our discussions with Jiti Gao. We thank Bryan Kelly and Seth Pruitt for sharing the data for the empirical application. All errors are our own.
  } 
  , Qingliang Fan\footnote{
    Corresponding author. ELB 903, The Chinese University of Hong Kong, Shatin, Hong Kong. E-mail: michaelqfan@cuhk.edu.hk.
  }\\
  Department of Economics, The Chinese University of Hong Kong
}

\maketitle

\begin{abstract}	
We propose a characteristics-augmented quantile factor (QCF) model, where unknown factor loading functions are linked to a large set of observed individual-level (e.g., bond- or stock-specific) covariates via a single-index projection. The single-index specification offers a parsimonious, interpretable, and statistically efficient way to nonparametrically characterize the time-varying loadings, while avoiding the curse of dimensionality in flexible nonparametric models. Using a three-step sieve estimation procedure, the QCF model demonstrates high in-sample and out-of-sample accuracy in simulations. We establish asymptotic properties for estimators of the latent factor, loading functions, and index parameters. In an empirical study, we analyze the dynamic distributional structure of U.S. corporate bond returns from 2003 to 2020. Our method outperforms the benchmark quantile Fama-French five-factor model and quantile latent factor model, particularly in the tails ($\tau=0.05, 0.95$). The model reveals state-dependent risk exposures driven by characteristics such as bond and equity volatility, coupon, and spread. Finally, we provide economic interpretations of the latent factors.

\end{abstract}

\noindent%
{\it Keywords:} Quantile latent factor, panel nonlinear regression, single-index model

\newpage


	\section{Introduction}\label{sec:intro}
	
Factor models are central to high-dimensional modeling across economics and statistics, offering a powerful framework for summarizing common variation in multivariate data. A prominent application is in empirical asset pricing, where latent factor structures are used to explain the cross-section of asset returns. Traditional linear factor models, such as those estimated via principal component analysis (PCA), provide dimension reduction but lack interpretability and flexibility, especially when factor loadings depend on observable firm characteristics.

Recent advances enhance interpretability through characteristics-based factor models. These studies model factor loadings as functions of observed characteristics \citep[e.g.,][]{connor2007semiparametric, fan2016projected}. However, these models typically assume (i) additivity, (ii) time-invariant loadings, and (iii) the existence of the fourth moment. Such restrictions are often unrealistic in financial data: characteristic effects may interact nonlinearly \citep{daniel1999market}, loadings (i.e., risk exposure) evolve over time \citep{gagliardini2020estimation}, and asset returns exhibit heavy tails \citep{ibragimov2015heavy}. This motivates the need for a more flexible factor modeling framework that accommodates nonlinear, time-varying loadings and captures the entire conditional distribution of returns via quantile regression.

This paper proposes a quantile characteristics-augmented factor (QCF) model that addresses all three limitations. Our model introduces a semiparametric single-index structure to flexibly capture nonlinear and interactive effects of observed characteristics on time-varying loadings. By employing quantile regression, the model also accommodates heavy tails and heteroskedasticity, allowing analysis of the entire distribution and not just conditional means \citep{koenker1978regression}.

This paper contributes to several strands of literature, including semiparametric factor models, quantile factor models, quantile single-index models, and empirical asset pricing. Below, we review the most relevant theoretical and empirical works to position our contribution.

A large literature studies semiparametric latent factor models, where factor loadings are modeled as functions of observable characteristics. This includes characteristics-based factor models \citep{connor2007semiparametric}, instrumental PCA \citep{kelly2019characteristics}, and recent developments such as sieve-based generalized additive model (GAM) structures: projected PCA \citep{fan2016projected}, state-varying semiparametric factor model \citep{su2024sieve}, to name a few. These models improve interpretability over traditional PCA, but they impose additive separability across characteristics to sidestep the curse of dimensionality \citep{stone1985additive}. Although computationally convenient, additivity rules out non-linear interactions that are empirically important - for instance, the momentum effect is stronger among growth stocks \citep{daniel1999market}. Some studies introduce interaction terms manually \citep{freyberger2020dissecting}, but this becomes impractical as dimensionality grows.
Recent works explore more flexible loading specifications using deep learning \citep{feng2023deep, fan2022structural}, but these methods are computationally intensive and lack formal inference guarantees.

Our approach adopts a semiparametric single-index structure for factor loadings. To our knowledge, this is the first model to connect latent factor loadings with many observable individual-specific characteristics through a single-index formulation. This structure not only captures nonlinear and interactive effects while mitigating the curse of dimensionality, but also admits a compelling connection to machine learning: a single-index model can be viewed as a shallow neural network with one hidden unit and an unspecified activation function. In this sense, our approach offers a low-complexity, interpretable alternative to deep learning models for characteristic-based loadings.

Another distinct feature of our model is the incorporation of time-varying loadings, addressing a key limitation in the existing semiparametric latent factor literature. Most prior work assumes factor loadings are functions of time-invariant characteristics \citep[e.g.,][]{connor2007semiparametric, fan2016projected, kim2021arbitrage}, limiting their ability to capture dynamic shifts in risk exposure. In contrast, we allow both firm characteristics and their effects on loadings to evolve over time, reflecting growing evidence that factor exposures are state-dependent and driven by both firm-level and macroeconomic variables \citep{ferson1996measuring, gagliardini2020estimation, chen2023semiparametric}.
	
Moreover, the aforementioned literature on latent factor models all concentrates on the mean relationship, while in this paper we study the whole distribution in a quantile model and in the empirical study the tail risk, for example, is of great interests to many. The literature on quantile factor models remained limited until very recently due to the technical challenges posed by the non-smooth quantile objective.
\cite{chen2021quantile} introduces a quantile factor model with latent factors. Other related works include additional observed covariates as regressors in the quantile factor model \citep{ando2020quantile, ando2022quantile, ando2023spatial,belloni2023high}.
The models that are most closely related to ours are \cite{ma2021estimation} and \cite{chen2023estimation}, which can be viewed as quantile extensions of \cite{connor2012efficient} and \cite{fan2016projected}, respectively. Both papers consider factor loadings as additive functions of low-dimensional time-invariant covariates, which is somewhat restrictive in empirical studies, such as empirical asset pricing, as discussed before. Our model overcomes these limitations by allowing loadings to be (i) nonadditive across asset-specific characteristics via a single index structure and (ii) time-varying, accommodating dynamic characteristics such as equity momentum.

For our QCF model, we propose a three-step procedure that leverages sieve methods to estimate the single-index loading functions. In particular, we adopt a Hermite polynomial expansion to approximate the unknown index functions, which avoids compact support assumptions and enables a linear representation for efficient computation. This allows us to bypass iterative or profile estimation, which is often computationally intensive and sensitive to initial values. Under appropriate identification conditions, we obtain consistent estimators for factors, loading indices, and index coefficients via penalized quantile regression and eigen-decomposition.

Our model offers both methodological innovation and empirical relevance. A growing literature explores the relationship between asset returns and a large set of firm characteristics using modern machine learning methods, such as factor-augmented regressions \citep{ludvigson2009macro} and deep neural networks \citep{gu2020empirical}. While these approaches are powerful for prediction, they often neglect the underlying economic structure, do not identify interpretable risk factors, and ignore the explanatory power of characteristics on explaining the firm-specific risk exposures. 
Our framework bridges the gap between structural factor models and machine learning-based return prediction by embedding firm characteristics into a structural factor framework. It captures nonlinear and time-varying effects across the entire return distribution, combining the flexibility of modern predictive methods with the economic interpretability of structural modeling.

While the factor model literature has traditionally focused on equities, our empirical application centers on corporate bonds, which present distinct challenges. Bond returns tend to exhibit heavier tails, are more sensitive to macroeconomic conditions, and often feature lower liquidity \citep{bessembinder2009measuring}. A growing literature documents complex factor structures in bond markets \citep{GunaydinHollifield2019}, and recent methods such as instrumental PCA \citep{kelly2023modeling} use firm characteristics to interpret latent risk exposures more effectively. Machine learning approaches further underscore the importance of capturing nonlinear interactions: for example, \citet{bell2024glass} highlight how combinations of bond-level variables, such as bond spreads, and financial uncertainty, jointly influence return predictability in non-additive ways. Our model bridges these strands by incorporating time-varying, single-index loading functions into a quantile factor framework. This unified approach captures heavy-tailed bond return distributions and complex covariate interactions while maintaining interpretability and structural coherence. 
	
To summarize, our contributions are threefold. Methodologically, we propose a novel characteristics-augmented quantile factor (QCF) model that flexibly captures non-separable, time-varying factor loadings and models the full distribution, extending beyond mean-based approaches. Theoretically, we establish the asymptotic properties of the estimators, including convergence rates and central limit theorems for the parameters of interest as both the cross-sectional and temporal dimensions diverge. Empirically, we apply the QCF model to U.S. corporate bond returns from 2003 to 2020. Our findings demonstrate that the model outperforms existing benchmarks, including the quantile Fama-French 5 factor model and the quantile (latent) factor model, by capturing nonlinear, quantile-dependent risk structures, particularly in the tails of the return distribution. Furthermore, we provide economic interpretations of the estimated latent factors and, through the index structure, offer insights into how bond characteristics influence factor exposures in a manner that varies across different market conditions. 

The rest of the paper is structured as follows. In Section~\ref{sec:method}, we introduce the QCF model and propose a three-step estimation procedure. In Section~\ref{sec:asymptotic}, we first provide the necessary model assumptions and then establish the asymptotic properties for the proposed factor, loading and index estimators. In Section~\ref{sec:simulation}, we assess the finite sample performance through Monte Carlo simulations. Section \ref{sec:empirical} studies the return distribution of U.S. corporate bonds, and Section~\ref{sec:conclusion} concludes the paper.
	
Before proceeding, we introduce some notations that are used throughout the paper.
Let $\| \cdot \|$ denote the Euclidean norm for vectors and the spectral norm for matrices, that is $\|a\| :=\sqrt{a'a}$ and $\|A\|:= \sup_{a \not = 0}\| A a \| /\| a \|$ for a column vector $a$ and a matrix $A$. 
Let $I_p$ denote the $p$-dimensional identity matrix, whose dimension varies according to the subscript. Let $\varrho_i(A)$ denote the $i$-th largest eigenvalue of matrix A, and specifically, the largest and smallest eigenvalues are denoted by $\varrho_{\max}$ and $\varrho_{\min}$ respectively.
Let $1\{\cdot\}$ denote the indicator function. We denote by 
$a \vee b:=\max \{a, b\}$ and $a \wedge b:=\min \{a, b\}$ for scalars $a, b$. 

In addition, we provide a basic introduction to the Hilbert space $L^2(\R, e^{-w^2/2}):= \{g(w), \allowbreak \int_{\R} g^2(w) e^{-w^2/2} dw < \infty \}$, which contains the loading functions considered in this paper. The inner product is defined by $\langle g_1, g_2 \rangle := \int_\R g_1(w)g_2(w)e^{-w^2/2}dw$ and the induced norm $\|g\| := \sqrt{\langle g, g \rangle}$. Two functions $g_1(w)$ and $g_2(w)$ are called orthogonal if $\langle g_1, g_2 \rangle = 0$. The theory about Hilbert space can be found in standard textbooks on functional analysis \citep{young1988introduction}.
\section{Methodology}\label{sec:method}
	In this section, we introduce the quantile characteristics-augmented factor (QCF) model and propose an estimation procedure based on the sieve approach.
	
	\subsection{Model}\label{sec:model}
	To solve the problems mentioned in Section \ref{sec:intro}, for a given quantile $\tau \in (0,1)$, we assume the conditional quantile of $y_{it}$ admits the following model:
	\begin{equation}\label{model}
		Q_{y_{it}}(\tau |x_{it},f_t) =
		\sum_{k=1}^{r}f_{t(k),\tau} \lambda_{k,\tau}(x_{it}'\theta_k) , \quad i = 1,...,N; t = 1,...,T, 
	\end{equation}
	where 
	$f_{t,\tau} \equiv (f_{t(1),\tau}, ..., f_{t{(r)},\tau})' \in \mathbb{R}^{r}$ is the unobserved common factor vector, heterogeneous over quantile $\tau \in (0,1)$, $\lambda_{\tau}(\cdot) \equiv [\lambda_{1,\tau}(\cdot), ...,\lambda_{r,\tau}(\cdot)]': \R \to \R^r$ is the corresponding $r$-dimensional unknown loading function.  $x_{it} \in \mathbb{R}^{d}$ are observable characteristics (covariates) and index parameters $\theta_k := [\theta_{k(1)},...\theta_{k(d)}]'$ are allowed to be heterogeneous across  $k=1,...,r$.\footnote{$\theta_k$ is also allowed to vary across quantiles, for notational simplicity, we drop the dependency in the subscript.} 
	We consider the dimension of characteristics $d$ and latent factors $r$ to be finite, while the dimension of cross-sections $N$ and the number of observed time units $T$ are allowed to diverge to infinity simultaneously. For notational simplicity, we henceforth drop the quantile-dependence $\tau$ in the subscript unless it causes confusion.

Our proposed Model~\ref{model} distinguishes itself from the existing literature on characteristic-based quantile factor models \citep{ma2021estimation, chen2023estimation} by adopting a novel single-index structure for the loading function, in contrast to the GAM specification commonly used in previous studies. While the standard GAM specification avoids the curse of dimensionality by modeling only the marginal effect of each covariate additively, it does not capture nonlinear interactions between covariates. In contrast, our single-index specification captures these nonlinear interactions in a parsimonious manner, without incurring the complexity and estimation challenges associated with nonparametrically modeling high-dimensional interaction terms. Additionally, our single-index structure can be interpreted as a structured neural network with a single hidden layer \citep{Vaughan2018xNN}, providing a low-complexity and interpretable alternative to deep learning models for characteristic-based asset pricing \citep{feng2023deep}.

Furthermore, our framework extends the flexibility of characteristic-based quantile factor models by allowing the factor loadings to depend on time-varying characteristics, which are frequently observed in asset pricing applications. In contrast, prior work such as \citet{ma2021estimation} and \citet{chen2023estimation} restricts the loading functions to depend only on time-invariant covariates. This extension enhances the empirical applicability of our model in dynamic environments where asset characteristics evolve over time.

The key parameter of interests are $\{f_t, t=1,...,T\}$, $\lambda(\cdot)$ and $\{\theta_k,k=1,...,r\}$. 
For the link functions, we consider $\lambda_{k}(w) \in L^2(\R, e^{-w^2/2})$. The considered Hilbert space is sufficiently large and contains at least all bounded functions, polynomials, power functions, and even some exponential functions.
Also, our approach relaxes the conventional assumption of compact support in the sieve literature \citep{newey1997convergence, ma2016inference}. We note that as long as one element of $x_{it}$ possesses unbounded support, the domain of $\lambda_{k}(\cdot)$ should be treated as unbounded. To work with the unbounded support, we employ the Hermite polynomial sequence and approximate the link function by a truncated series.
	
	The Hermite polynomial sequence $\{h_\ell(\cdot), \ell = 0, 1, ...\}$, defined as
	\begin{align*}
		h_{\ell}(w) = \frac{(-1)^{\ell} }{\sqrt{\ell !}}\cdot e^{w^2/2} \cdot \frac{d^{\ell}}{d w^{\ell}} e^{-w^2/2}, \quad \ell=0,1,...
	\end{align*}
	forms an orthogonal basis of $L^2(\R, e^{-w^2/2})$, so that for any $ g(w) \in L^2(\R, e^{-w^2/2})$, we can expand it in terms of $h_\ell(w)$ as $g(w) = \sum_{m=0}^{\infty} b_\ell h_{\ell}(w)$, where $b_\ell = (2\pi)^{-1/2} \langle g, h_\ell \rangle$.
	With the above notations, we expand
	\begin{align}\label{eq1}
		\lambda_{k}(x_{it}'\theta_k) = \sum_{\ell=0}^{m-1}b_{k,\ell} h_\ell(x_{it}'\theta_k) + \Delta_{k,m}(x_{it}'\theta_k), \quad 
		k = 1,...,r
	\end{align}
	where $b_{k,\ell} \equiv b_{k,\ell,\tau} := (2\pi)^{-1/2} \langle \lambda_{k,
    \tau}, h_\ell \rangle$ is quantile-dependent, $m$ is the truncation parameter and $\Delta_{k,m}(x_{it}'\theta_k)$ is the truncation residual which is negligible (see Assumption~\ref{ass.cov}.(iii)).
	
	Given the above expansion, one can estimate $\theta_k$ and $\lambda_k(\cdot)$ with a profile method or an iterative estimation method \citep{ma2016inference}. However, such a method is computationally heavy and unstable in practice, as the numerical estimates are often sensitive to the initial values due to the multimodality or flatness of a curve. To address this issue, we leverage the property of the Hermite polynomials (see Lemma 12.4.2 of \cite{blower2009random}) to further write each term in $	\lambda_{k}(x_{it}'\theta_k)$ as
	\begin{align}\label{eq2}
		b_{k,\ell} h_\ell(x_{it}'\theta_k)
		= \sum_{|\mathbf{p}| = \ell} \gamma_{k,\ell,\mathbf{p}}(\theta_k) \mathcal{H}_{\mathbf{p}}(x_{it}), 
		\quad \ell =1,2, ...,
	\end{align} 
	where 
	$\mathbf{p} := (p_1,...,p_d)'$ with nonnegative integer $p_j$ for $j=1...,d$ and $|\mathbf{p}| := \sum_{j=1}^{d}p_j$,
	\begin{align}\label{eq3}
		\gamma_{k,\ell, \mathbf{p}}(\theta_k) := 
		\sqrt{\frac{\ell !}{p_1! ... p_d!}} b_{k,\ell} \prod_{j=1}^{d}\theta_{k(j)}^{p_j},
		\quad
		\mathcal{H}_{\mathbf{p}}(x_{it})
		:= \prod_{j=1}^{d}h_{p_j}(x_{it(j)}),
	\end{align}
	where $x_{it(j)}$ is the $j$th element of $x_{it}$. Let $\mathcal{D}_\ell := \{\mathbf{p}: |\mathbf{p}| = \ell\}$, and 
	$\mathcal{D} := \bigcup_{\ell=0}^{m-1} \mathcal{D}_\ell$. It is easy to show that $\mathcal{D}_\ell$ contains $\binom {d+\ell-1}{d-1}$ elements and $\mathcal{D}$ contains $M :=  \sum_{\ell=0}^{m-1} \binom {d+\ell-1}{d-1} =  \binom {m+d-1} {d} = O(m^{d})$ elements.
    
	Then, (\ref{eq1})-(\ref{eq3}) implies that the single-index factor loading function in \eqref{model} can be factorized into an additive form:      
	\begin{align*}
		\lambda_{k}(x_{it}'\theta_k) = \gamma_{k,m}'H_{m}(x_{it}) + \Delta_{k,m}(x_{it}'\theta_k),
	\end{align*}
	where $H_m(\cdot): \R^d \to \R^M$, is a $M \times 1$ column vector of functions consisting of all terms $\mathcal{H}_{\mathbf{p}}(\cdot)$ for all $\mathbf{p} \in \mathcal{D}$ and $\gamma_{k,m}:= [\gamma_{k,m(1)},...,\gamma_{k,m(M)}]'$ consists of all coefficients $ \gamma_{k,\ell, \mathbf{p}}(\theta_k)$ defined conformably, termed as the intermediate loading coefficient in what follows.  
	
	Thus, the model (\ref{model}) can be re-written as 
	\begin{align} \label{eq.Qy_expansion}
		Q_{y_{it}}(\tau |x_{it},f_t)
		=
		f_{t}' \Gamma_{m}' H_{it} + \eta_{it}  , \quad i = 1,...,N; t = 1,...,T,
	\end{align}
	where $\Gamma_{m} := [\gamma_{1,m},...,\gamma_{r,m}]$, $H_{it} \equiv H_m(x_{it})$, and $\eta_{it} :=\sum_{k=1}^{r} f_{t(k)} \Delta_{k,m}(x_{it}'\theta_k)$ is a negligible approximation error, regularized by Assumption~\ref{ass.cov}.(iii).

\subsection{Model Identification}\label{sec:identification}
    For identification purposes, there are two sets of identification restrictions that need to be imposed on QCF model. The first is imposed on the factor structure to separately identify $\{f_t\}$ and $\{\lambda_k(x'\theta_k),k=1,...,r\}$, while the other is imposed to uniquely identify the index parameters. Assumption~\ref{ass.identification} formally states the sufficient identification conditions.

    \begin{assumption} \label{ass.identification}
		\begin{enumerate}[(i)]
			\item $T^{-1}F'F = I_r$ and $ 
			\Gamma_m' \Gamma_m$ is a $r \times r$ diagonal matrix with distinct entries ($r \leq M$).
            \item $\|\theta_k^0\| = 1$ with its first element $\theta_{k(1)}^0 >0$ for all $k = 1,...,r$.
		\end{enumerate}
	\end{assumption}

   Assumption~\ref{ass.identification}.(i) imposes $r^2$ restrictions, sufficient to identify the latent factor $F$ and intermediate loading coefficient $\Gamma_m$ in model (\ref{eq.Qy_expansion}). The matrix factorization $\Psi := [\psi_1,...,\psi_T]' =  F \Gamma_m' $ requires $r^2$ restrictions for unique identification  \citep{bai2003inferential,bai2013principal}.   
   Then, the identification of  $F$ and $\Gamma_m$ from model (\ref{eq.Qy_expansion}) further
   leads to the unique identification of $F$ and $\{\lambda_k(x_{it}'\theta_k), k=1,...,r \}$. 
   
 We note that the conventional restrictions in factor model literature \citep[e.g.,][]{bai2003inferential, bai2009panel} only require $\Gamma_m'\Gamma_m$ to be diagonal. Then, the factor and intermediate loading coefficient can be consistently estimated up to a rotation matrix, which depends on the estimator itself, and a column sign change, see Theorem~\ref{t2:F}. The rotation indeterminacy is a well-known issue in the factor model literature. It not only raises the question of what does the latent factor consistently estimates, but also further impedes the recovery and interpretation of the estimated index parameter $\theta_k$ in the QCF model. To ease this issue, we further impose the restriction that the diagonal elements of $\Gamma_m'\Gamma_m$ are distinct, motivated by the PC1 condition of \cite{bai2003inferential}. We show in Theorem~\ref{lemma.A3}.(i) that the rotation matrix is an identity matrix asymptotically, such that the latent factor, intermediate loading coefficient, and consequently the index parameter, can be consistently estimated without rotation asymptotically.

In addition, to link the identification restriction of the intermediate loading coefficient $\Gamma_m$ with $\lambda(\cdot)$ and $\{\theta_k\}$, we provide a few simple scenarios when Assumption~\ref{ass.identification}.(i) is satisfied in Appendix \ref{appendix:identification},

Assumption~\ref{ass.identification}.(ii) is a standard identification condition in literature of single-index models \citep[e.g.,][]{ichimura1993semiparametric}. Under this restriction, the relationship between $\gamma_{k,m} $ and $\gamma_{k,\ell, \mathbf{p}}(\theta_k)$, as well as $\gamma_{k,m}$ and $\{b_{k,\ell},\ell=0,...,m-1\}$, are mutually determined uniquely, as shown in \cite{dong2024semiparametric}. Thus, both the index parameter $\theta_k$ and the sieve coefficient $\{b_{k,\ell}, \ell=0,...,m-1\}$, and thus the loading function $\lambda_k(\cdot)$, can be recovered from $\gamma_{k,m}$ uniquely.

To begin with, we first uncover the relationship between $\gamma_{k,m}$ and $b_{k,\ell}$.  Define the subvector of $\gamma_{k,m}$ associated with $b_{k,\ell}$ as $\text{Sub}(\gamma_{k,m},\ell) := (\gamma_{k,\ell, \mathbf{p}}(\theta_k), |\mathbf{p}|=\ell)'$ for $\ell=0,...,k-1$. For $\ell=0$, it is straightforward that $b_{k,0} = \gamma_{k,m(0)}$. For $\ell = 1,..., m-1$, we pick the elements in $\text{Sub}(\gamma_{k,m},\ell)$ that have the form of the $\ell$-th power. They are 
	\begin{align}\label{sub_gamma}
		\gamma_{k,m(\ell_1)} = b_{k,\ell}\theta_{k,1}^{\ell}, ...,\ \gamma_{k,m(\ell_d)} = b_{k,\ell}\theta_{k,d}^{\ell},
	\end{align}
	where the corresponding subindexes are $\ell_1 := \frac{(d+\ell-1)!}{d! (\ell-1)!}+1, \ell_2 := \ell_1 + \frac{(d+\ell-1)!}{(d-1)!\ell!}-\frac{(d+\ell-2)!}{(d-2)!\ell!},..., \ell_d := \frac{(d+\ell)!}{d! \ell!}$. The relationship in (\ref{sub_gamma}) suggests that, under the identification condition $\theta_{k(1)} > 0$, 
	\begin{align}\label{b_ell}
		b_{k,\ell} = \text{sgn}(\gamma_{k,m(\ell_1)})\bigg(\sum_{j=1}^d \big(\gamma_{k,m(\ell_j)}\big)^{2/\ell}\bigg)^{\ell/2}.
	\end{align}
	Next, we recover $\theta_{k}$ from $\gamma_{k,m}$. For $\ell = 1,..., m-1$, let $Q_\ell := [0, I_d,0]$ be a $d \times M $ matrix, where the first zero matrix has dimension $d \times \frac{(d+\ell-1)!}{d! (\ell-1)!}$, while the second zero matrix has a conformable dimension according to $\Gamma_m$. Then, $Q_{\ell}$ is the selection matrix to pick up the first $d$ elements from $\text{Sub}(\gamma_{k,m},\ell) $, that is, 
	$$Q_{\ell}\gamma_{k,m} = (b_{k,\ell} \theta_{k(1)}^{\ell},\sqrt{\ell} b_{k,\ell} \theta_{k(1)}^{\ell-1}\theta_{k(2)},...,\sqrt{\ell} b_{k,\ell} \theta_{k(1)}^{\ell-1}\theta_{k(d)} )',$$
	which implies that, if $b_{k,\ell} \neq 0$,
	\begin{align}\label{theta_recover}
		\theta_{k} = D_{k,\ell}Q_{\ell}\gamma_{k,m}, 
		\quad 
		\text{where } D_{k,\ell} := \text{diag}\bigg( \frac{1}{b_{k,\ell} \theta_{k(1)}^{\ell-1}},  \frac{1}{\sqrt{\ell} b_{k,\ell} \theta_{k(1)}^{\ell-1}}, ...,\frac{1}{\sqrt{\ell} b_{k,\ell} \theta_{k(1)}^{\ell-1}} \bigg),
	\end{align}
	where $\theta_{k(1)} = (\gamma_{k,m(\ell_1)}/ b_{k,\ell_1})^{1/\ell}$ given the relationship in (\ref{b_ell}).

\subsection{Estimation}\label{sec:est}
Given the above sieve approximation (\ref{eq.Qy_expansion}), we propose a three-step quantile estimation approach. First, we estimate $\widehat{\Psi} :=[\widehat{\psi}_1,...,\widehat{\psi}_T]'$ via a penalized quantile regression. Second, we uniquely identify $\widehat{F}:= [\widehat{f}_1,...,\widehat{f}_T]'$ and $\widehat{\Gamma}_m:= [\widehat{\gamma}_{1,m},...,\widehat{\gamma}_{r,m}]$ via eigen-decomposition according to the factor identification conditions. Finally, we recover $\{\widehat{\theta}_k\}$ and $\{\widehat{\lambda}_k(\cdot)\}$ from $\widehat{\gamma}_{k,m}$ given the mutually determined relationships among the parameters in the single-index specification for each $k=1,...,r$, respectively. Algorithm~\ref{algorithm} summarizes our estimation procedure.  

We note that, although the observed covariates are finite-dimensional, $\Psi$ is high-dimensional with an order of $O(m^d)$, and exhibits strong correlations due to its underlying structure. To avoid overfitting and enhance estimation stability, we adopt a ridge penalty in the quantile estimation, which has been shown to be effective in the simulations in Section~\ref{sec:simulation.1}.  Ridge regularization is particularly well-suited to our setting, as it does not rely on sparsity assumptions and instead provides continuous shrinkage that accommodates correlated and moderately high-dimensional parameters.
Other types of penalty function can also be considered. For example, LASSO-type shrinkage can be adopted if we consider a high-dimensional characteristics setting. It requires additional assumptions, such as sparsity structure in the link function $\lambda_{k}(\cdot)$ via the coefficients $b_{k,\ell}$, or in the index parameter $\theta_k$, as discussed in, for example, \cite{dong2024semiparametric}. The high-dimensional characteristics setting is beyond the scope of the present study.

\begin{algorithm}
\caption{QCF Three-step Estimation\label{algorithm}}
\begin{algorithmic}
\State \textbf{Step 1:} Estimate $\widehat{\Psi}$ via ridge-regularized quantile regression:
\[
\widehat{\psi}_t := \underset{\psi_t}{\text{arg min}}\frac{1}{NT}\sum_{t=1}^T\sum_{i=1}^N
\rho_\tau(y_{it} -\psi_t' H_{it}) + \frac{a_{NT}}{2}\sum_{t=1}^{T}\|\psi_t\|^2,
\]
where $\rho_\tau(u) = u(\tau - 1\{u <0\})$ and $a_{NT} >0$ is a penalty parameter.

\State \textbf{Step 2:} Identify $\widehat{F}$ and $\widehat{\Gamma}_m$ given factor identification restrictions (Assumption~\ref{ass.identification}.(i)) via eigen-decomposition:
\[
\frac{1}{T}\widehat{\Psi}\widehat{\Psi}'\widehat{F} = \widehat{F}\widehat{V},
\]
where $\widehat{V}$ is the $r \times r$ diagonal matrix of the first $r$ largest eigenvalues of $T^{-1}\widehat{\Psi}\widehat{\Psi}'$ in descending order, and $\widehat{F}$ is $\sqrt{T}$ times eigenvectors of the first $r$ largest eigenvalues of $T^{-1}\widehat{\Psi}\widehat{\Psi}'$. Then, we estimate
$ \widehat{\Gamma}_m := T^{-1}\widehat{\Psi}'\widehat{F}$.

\State \textbf{Step 3:} For $k=1,...,r$, recover $\widehat{\theta}_k$ and $\widehat{\lambda}_k(\cdot)$ under index identification restriction (Assumption~\ref{ass.identification}.(ii)):
\begin{itemize}
    \item 
  Recover $\widehat{\theta}_k$ is obtained using the sample analog of (\ref{b_ell})-(\ref{theta_recover}):
\[
\widehat{\theta}_k :=  \widehat{D}_{k,\ell}Q_{\ell}\widehat{\gamma}_{k,m},
\]
where
\begin{align}\label{D_hat}
  \widehat{D}_{k,\ell} := \text{diag}\left( \frac{1}{\widehat{b}_{k,\ell} \widehat{\theta}_{k(1)}^{\ell-1}},  \frac{1}{\sqrt{\ell} \widehat{b}_{k,\ell} \widehat{\theta}_{k(1)}^{\ell-1}}, \dots, \frac{1}{\sqrt{\ell} \widehat{b}_{k,\ell} \widehat{\theta}_{k(1)}^{\ell-1}} \right),
\end{align}
and
\[
\widehat{b}_{k,\ell} := \text{sgn}(\widehat{\gamma}_{k,m(\ell_1)})\left(\sum_{j=1}^d \left(\widehat{\gamma}_{k,m(\ell_j)}\right)^{2/\ell}\right)^{\ell/2},
\]
\[
\widehat{\theta}_{k(1)} := \left(\frac{\widehat{\gamma}_{k,m(\ell_1)}}{\widehat{b}_{k,\ell_1}}\right)^{1/\ell}, \quad \text{provided that } \widehat{b}_{k,\ell}\widehat{\theta}_{k(\ell)} \neq 0.
\]

\item Estimate $\widehat{\lambda}_{k}(\cdot) := \sum_{\ell=0}^{m-1}\widehat{b}_{k,\ell} h_\ell(\cdot)$, where
\begin{align*}
(\widehat{b}_{k,m},\  \ell=0,...,m-1;\ & k=1,...,r) \\ 
:=\ & \underset{b_{k,\ell}}{ \text{arg min}} \frac{1}{NT}
\sum_{t=1}^T \sum_{i=1}^N
\rho_\tau \Big(y_{it} - \sum_{k=1}^{r}\sum_{\ell=0}^{m-1}\widehat{f}_{t(k)} h_\ell(x_{it}'\widehat{\theta}_k)b_{k,\ell}  \Big).
\end{align*}
\end{itemize}
\end{algorithmic}
\end{algorithm}

    \subsection{Practical Implementation}
This section illustrates the implementation of the proposed QCF three-step estimation procedure for practitioners.

{\bf Selection of tuning parameters.} In practice, the number of factors $r$ and approximated sieve basis functions $m$ in QCF model are unknown, and the penalty parameter $a_{NT}$ needs to be given by users. In our simulation study and empirical analysis, we choose $r$, $m$ and $a_{NT}$ simultaneously through a data-driven selection method given as \begin{align}\label{eq:para_select}
		(\widehat{r}, \widehat{m}, \widehat{a}_{opt}) = \underset{r,m,a_{NT}}{\text{arg min}} \frac{1}{N}\sum_{i=1}^N \rho_\tau \big(y_{iT} - \sum_{k=1}^{r} \widehat{f}_{T(k)} \widehat{\lambda}_k(x_{iT}'\widehat{\theta}_k)\big),
	\end{align}
	where $\widehat{\lambda}_k(\cdot)$ and $\widehat{\theta}_k$ are estimated with the above algorithm using data from $t=1$ up to $T-1$. Then, $\widehat{f}_T$ is estimated via quantile regression \citep{koenker1978regression} of $y_{iT}$ on $\{\widehat{\lambda}_k(x_{iT}'\widehat{\theta}_k),k=1,...,r\}$ for $i=1,...,N$. We set $a_{NT} = 0$ and select $(\widehat{r}, \widehat{m})$ if considering QCF estimation without penalty.   
    We note that the latent factor $f_T$ cannot be predicted given the data from $t=1$ to $T-1$ without imposing additional structure on $\{f_t, t=1,...,T-1\}$. Strictly, 
	$\widehat{f}_T$ is not a one-step-ahead forecast of the latent factor, but rather an out-of-sample estimation with the cross-sectional regression, which is of empirical interest in asset pricing literature \citep[e.g.,][]{kelly2019characteristics,kelly2023modeling,feng2024deep}.

   {\bf Unbalanced panel.} Another practical issue that often arises in financial application is that, in the time period $t$, we only observe $n_t$ observations. In this case, we replace $\sum_{t=1}^T\sum_{i=1}^N$ with $\sum_{t=1}^T\sum_{i=1}^{n_t}$ and $NT$ with $\sum_{t=1}^T n_t$ in Steps 1 and 3 of Algorithm \ref{algorithm}, while the remaining steps proceed as usual.
	
\section{Asymptotic Theory}\label{sec:asymptotic}
In this section, we study the asymptotic properties of the key estimators. We use superscript $0$ to distinguish the true parameters in what follows.
	
\subsection{Assumptions}
We first provide the necessary assumptions to establish the convergence rates and asymptotic distributions. 
    
\begin{assumption}\label{ass.cov}
		\begin{enumerate}[(i)] For all $i=1,...,N$ and $t=1,...,T$, we have
			\item there exist $\delta >0$, such that $\mathbb{E}[\|f_{t}^0\|^{4+\delta}]  < \infty$.
			\item Let $\Sigma_{H_t} :=N^{-1}\sum_{i=1}^N \mathbb{E}[H_{it}H_{it}']$. There exist $c_1$ and $c_2$ such that $0 < c_1 \leq \varrho_{min}(\Sigma_{H_t}) \leq \varrho_{\max}(\Sigma_{H_t}) \leq c_2 < \infty$ for all $t=1,...,T$.
			\item There exists $\delta_0 > 0$ such that $\sup_{0 \leq \varepsilon \leq 1} \sup_{\|\theta_k - \theta_k^0\| \leq \varepsilon}\max_{1 \leq i \leq N, 1 \leq t \leq T} \E\big[\big(\Delta_{k,m}^0(x_{it}'\theta_k^0)\big)^2\big] = O(m^{-\delta_0})$ for all $k =1,...., r$. 
			\item $ \max_{1 \leq i \leq N, 1 \leq t \leq T} \E\big[\|\mathcal{H}_{\mathbf{p}}(x_{it})\|^{4 + \delta}\big] = O(|\mathbf{p}|^d)$ as $|\mathbf{p}| \to \infty$, where $|\mathbf{p}| $ is given under (\ref{eq2}).
		\end{enumerate}
	\end{assumption}
	
	The conditions in Assumption \ref{ass.cov} are all standard in the literature on factor models and sieve estimations. In particular
	Assumptions~\ref{ass.cov}.(ii)-(iv) are in the sprite of Assumption 3.(i)-(ii) of \cite{dong2015semiparametric}.

	\begin{assumption}\label{ass.error}
		Define the filtration $\mathcal{F}_t := \sigma(\epsilon_{s-1}, X_{s}, f_{s}^{0}, s \leq t+1)$, where $\epsilon_{t}:= (\epsilon_{1t},...,\epsilon_{Nt})'$ and $X_{t}:= (x_{1t},...,x_{Nt})'$. We assume 
		\begin{enumerate}[(i)]
			\item $\{\epsilon_{it},i=1,...,N \}$ are mutually independent conditional on $\mathcal{F}_{t-1}$ for each $t \geq 1$.
			
			\item $\varphi_\tau(\epsilon_{t})$ forms a martingale difference sequence (MDS), that is
			$\mathbb{E}[\varphi_\tau(\epsilon_{t}) | \mathcal{F}_{t-1}] = 0$ almost surely, where $\varphi_\tau(u) := 1(u < 0) - \tau$ is the subgradient of the quantile loss function $\rho_\tau(\cdot)$.
			
			\item The conditional density function of $\epsilon_{it}$ given $\mathcal{F}_{t-1}$, denoted as $g_{it}(\cdot)$, is Lipschitz continuous of order 1, and uniformly bounded away from 0 almost surely, and $g_{it}(0) = \tau$ almost surely.
		\end{enumerate}
	\end{assumption}
	
	Assumption~\ref{ass.error} allows for the heteroscedastic and serially-correlated errors, which relaxed the assumption of conditional independence in the literature of quantile factor models \citep{ando2020quantile,chen2021quantile}.Condition~\ref{ass.error}.(ii) automatically holds if $\{\epsilon_{t}\}$ are independently distributed. Condition~\ref{ass.error}.(iii) is a technical requirement in related calculations. Specifically, $g_{it}(\cdot)$ is required to be Lipschitz continuous such that $g_{it}(0)$ can be estimated by a kernel estimator; see Corollary~\ref{t6:Xi_estimates}.

	\begin{assumption} \label{ass.theta}
		For all $k = 1,...,r$,  $\theta_k^0 \in \Theta_k \subset \R^d$, where $\Theta$ is a convex and compact set and $\theta_k^0$ is an interior point of $\Theta_k$. For each $\theta_k  \in \Theta_k$, $x_{it}'\theta_k$ has density $g_{k}(w) $ such that $\sup_{\theta_k \in \Theta_k}g_k(w) \leq c e^{-w^2/2}$ for all large $|w|$ and some constant $c >0$.
	\end{assumption}
	
Assumption~\ref{ass.theta} is standard in the single-index model literature \citep[e.g.,][]{dong2015semiparametric, dong2024semiparametric}, and is primarily motivated by the use of sieve methods based on Hermite polynomial bases for estimating the 
unknown link function $\lambda(\cdot)$. This assumption effectively requires that the index $x_{it}'\theta_k$
  has tails no heavier than those of the normal distribution. Consequently, it accommodates many commonly used covariates, particularly those supported on compact sets or approximately normally distributed. However, this assumption excludes settings in which the index may exhibit heavy tails, a situation that can arise in financial applications. While addressing such cases lies beyond the scope of the present study, developing robust estimation techniques for heavy-tailed covariate indices is an important direction for future research.

	\subsection{Asymptotic Properties}
	
	In this section, we establish the rate of convergence and the central limit theorem (CLT) for the parameter of interests. We first derive the rates of convergence of the estimated factors $\widehat{f}_t$ and intermediate loading coefficients $\widehat{\Gamma}_m$ in Theorem~\ref{t2:F}.
	\begin{theorem}\label{t2:F}
		Suppose Assumptions~\ref{ass.identification}, \ref{ass.cov}-\ref{ass.theta} hold. Additionally, suppose (i) $a_{NT} = o(m^{3d/4}N^{-1/2})$, (ii)  $m^{3d/4}N^{-1/2} = o(1)$ and  (iii) $m^{3d/4}T^{-1/2} = o(1)$. Then, as $(N,T) \to (\infty,\infty)$, we have
		\begin{enumerate}[(i)]
			\item $\big\|\widehat{F} - F^0 R\big\|^2 =\sum_{t=1}^T \big\|\widehat{f}_t - R' f_t^0 \big\|^2 = O_P\big(\frac{T}{N} \vee  \frac{m^{3d/2}}{N} \vee T C_R^2\big)$;
			\item $\big\|\widehat{\Gamma}_m - \Gamma_m^0 (R^{-1})' \big\|^2 
			= O_P( \frac{m^{3d/2}}{NT} \vee \frac{m^{3d/2}}{N^2} \vee \frac{m^{3d/2}C_R^2}{N} \vee C_R^4 \vee \frac{C_R^2}{T} )$.
		\end{enumerate}
		where the rotation matrix $R := (\Gamma_m^{0\prime}\Gamma_{m}^0)(F^{0\prime}\widehat{F}/T)\widehat{V}^{-1}$ and $\widehat{V}$ is the $r \times r$ diagonal matrix of the first $r$ largest eignevalues of $\widehat{\Psi}\widehat{\Psi}'/T$ in the desending order, and $C_R = O( a_{NT} \vee a_{NT}^2 m^{3d/2}N^{-1} \ln(T) \vee m^{3d}N^{-1} \vee m^{ - \delta_0/2} \vee m^{9d/4}N^{-3/4}\sqrt{\ln (NT)})$.

	\end{theorem}
	
	The convergence rates for both $\widehat{F}$ and $\widehat{\Gamma}_m$ are jointly determined by the following three terms: (i) the divergence ratio between $N$ and $T$, which is common in factor analysis \citep[e.g.,][]{bai2003inferential}, (ii) the convergence rate of the nonparametric estimation, and (iii) the higher-order terms of the first step estimation $\widehat{\psi}_t$. Taking the results of Theorem~\ref{t2:F}.(i) as an example, the second term of the rate is associated with the single index approximation, similar to Lemma 3.1 of \cite{dong2015semiparametric}. Noting that $M = O(m^d)$, the rate is slower than the standard convergence rate of the series approximation $O_p(M/N)$ due to the fact that we cannot bound the basis functions uniformly. To solve this type of problem, Assumption 3.(iv) of \cite{su2012sieve} imposes a uniform bound on the $(4+\delta)$th-order moments of the basis functions. The terms involving $C_R$ are associated with the higher-order terms in the first-step estimator $\widehat{\psi}_t$, where $C_R$ is the uniform rate of the higher-order terms in the Bahadur representation of $\widehat{\psi}_t$ (see Proposition~\ref{t1:bahadur} in Appendix~\ref{appendix:proofA}).

	In the following theorem, we refine the rates by further regulating relationships among the penalty parameter $a_{NT}$, the number of sieve bases $m$, time $T$, and cross sections $N$. 
	\begin{theorem}\label{lemma.A3}
		Suppose Assumptions~\ref{ass.identification}, \ref{ass.cov}-\ref{ass.theta} hold, and $m \asymp N^{\delta_1}$, $T \asymp N^{\delta_2}$ and $a_{NT} \asymp N^{\delta_3}$ with $\delta_0 < \delta_1 < \frac{1}{9d}$, $0 < \delta_2 \leq 1$ and $\delta_3 < -\frac{1}{2}$. Then,
		\begin{enumerate}[(i)]
			\item $\big\|R - I_r \big\|^2 = O_P\big(\frac{1}{NT}\big)$, and $\|\widehat{V} - \Gamma_{m}^{0\prime}\Gamma_{m}^0\|^2  = O_P(\frac{1}{NT})$;
			\item $\big\|\widehat{f}_t - f_t^0 \big\|^2 = O_P\big(\frac{1}{T}\big)$ for a given $t=1,...,T$;
			\item $\big\|\widehat{\Gamma}_m - \Gamma_m^0 \big\|^2 = O_P\big(\frac{m^{3d/2}}{NT}\big)$;
			\item $\big\|\widehat{\theta}_k - \theta_k^0 \big\|^2 = O_P\big(\frac{1}{NT}\big)$ for a given $k=1,...,r$.
		\end{enumerate}
	\end{theorem}
	
	Theorem \ref{lemma.A3}.(i) suggests that the rotation matrix $R$ is asymptotically an identity matrix with an order of $\sqrt{NT}$; this rate is identical to the one under similar factor identification constraints PC1 in \cite{bai2013principal}. Then, we are able to show that $\widehat{f}_t$ converges to its true value at a standard convergence rate in the factor analysis by regulating the diverging rate of $m$. The convergence rate of the intermediate loading estimator $\widehat{\Gamma}_m$ is somewhat different from the literature because the factor loadings in QCF are homogeneous over $i$ and admit functional specifications. Indeed, it is easy to check that $\widehat{\Gamma}_m$ and $\widehat{\theta}_k$ achieve nonparametric rates comparable to the case where the factor $f_t^0$ is observed.

	In the following theorems, we establish the CLT for the estimators of factor, and loading function and index parameter.
	\begin{theorem}\label{t4:ft_CLT}
		Under the Assumptions of Theorem~\ref{lemma.A3}, as $(N,T) \to (\infty,\infty)$,
		\begin{align*}
			\Xi_{f_t,\tau}^{-1/2} \sqrt{N}(\widehat{f}_t - f_t^0) \overset{d}{\to} N(0,I_r),
		\end{align*} 
		where
		$$\Xi_{f_t,\tau} :=\underset{N \to \infty}{\text{plim}}\ \tau(1-\tau)(\Gamma_{m}^{0\prime}\Gamma_{m}^0)^{-1}\Gamma_{m}^{0\prime}\Sigma_{\epsilon H_t}^{-1}\Sigma_{H_t}\Sigma_{\epsilon H_t}^{-1}\Gamma_{m}^{0}(\Gamma_{m}^{0\prime}\Gamma_{m}^0)^{-1}$$		with $\Sigma_{H_t} := N^{-1}\sum_{i=1}^N \E[H_{it}H_{it}']$ and $\Sigma_{\epsilon H_t}$ is given in Theorem~\ref{t3:theta}.
	\end{theorem}

Theorem~\ref{t4:ft_CLT} establishes the CLT for $\widehat{f}_t$. We note that $\widehat{f}_t$ exhibits the usual $\sqrt{N}$-rate of consistency and asymptotic normality. 

	\begin{theorem}\label{t5:lambda_CLT}
    Denote $\Lambda^0(x,\theta^0) := [\lambda_1^0(x'\theta_1^0),...,\lambda_r^0(x'\theta_r^0)]'$ 
		, $\widehat{\Lambda}(x,\widehat{\theta}) := [\widehat{\lambda}_1(x'\widehat{\theta}_1),...,\widehat{\lambda}_r(x'\widehat{\theta}_r)]'$. 
		Under the Assumptions of Theorem~\ref{lemma.A3}, for $x \in \R^d$, we have
		\begin{align*}
	\Xi_{\lambda,\tau}^{-1/2} \sqrt{\frac{NT}{M^{3/2}}} \Big(\widehat{\Lambda}(x,\widehat{\theta}) - \Lambda^0(x,\theta^0)\Big) \overset{d}{\to} N(0,I_r),
		\end{align*}
		as $(N,T) \to (\infty,\infty)$, where 
		$$\Xi_{\lambda,\tau} :=\underset{{N,T \to\infty}}{ \text{plim}}\frac{ \tau(1-\tau)}{NTM^{3/2}} \sum_{i=1}^N\sum_{t=1}^T f_t^{0} H_{it}' \Sigma_{\epsilon H_t}^{-1}H_m(x)H_m(x)'\Sigma_{\epsilon H_t}^{-1}H_{it} f_t^{0\prime}.$$
	\end{theorem}

The above theorem establishes 
 the asymptotic normality of $\widehat{\Lambda}(x,\widehat{\theta})$, rather than $\widehat{\lambda}(w):= [\widehat{\lambda}_1(w),...,\widehat{\lambda}_r(w)]'$ itself in isolation. Inference on $\widehat{\Lambda}(x,\widehat{\theta})$
 remains practically relevant, as it captures the actual loadings used in estimating conditional quantile functions. 
 Inference at the index level remains both meaningful and sufficient for applied purposes, even if the global properties of $\widehat{\lambda}(w)$ are not fully characterized.
 
 Importantly, our focus is on inference for 
$\widehat{\theta}_k$, the index parameter governing the influence of observed characteristics on factor loadings, which is of primary empirical interest. Below, we establish the asymptotic normality for $\widehat{\theta}_k$.
    
    \begin{theorem}\label{t3:theta}
		Under the assumptions of Theorem~\ref{lemma.A3},
		as $(N,T) \to (\infty,\infty)$, 
          \begin{align*}
            \sqrt{NT} & \Big(\widehat{\theta}_{k} -\theta_k^0 \Big)
            \overset{d}{\to} N(0,\Xi_{\theta,\tau}),
        \end{align*}
        where 
        \begin{align*}
            \Xi_{\theta,\tau} :=  \underset{N,T \to \infty}{\text{plim}}\  \frac{\tau(1-\tau)}{NT} 
            \sum_{i=1}^N\sum_{t=1}^T
             & D_{k,\ell} Q_{\ell}  \Big\{  
             \Sigma_{\epsilon H_t}^{-1} H_{it} f_t^{0\prime} 
             + 
             2^{-1} \Gamma_m^{0}(\zeta_{it} + \zeta_{it}')
             \Big\} 
             u_k \\
             & \cdot u_k'
              \Big\{  f_t^0 H_{it}'
             \Sigma_{\epsilon H_t}^{-1} 
             + 
             2^{-1} (\zeta_{it} + \zeta_{it}')\Gamma_m^{0\prime}
             \Big\}  Q_{\ell}' D_{k,\ell}' ,         
        \end{align*}
        with $D_{k,\ell}$ and $Q_{\ell}$ defined Section~\ref{sec:model}, $\Sigma_{\epsilon H_t} : = \text{plim}_{N \to \infty}N^{-1}\sum_{i=1}^N H_{it}H_{it}'g_{it}(0)$,
		and $\zeta_{it} := (\Gamma_m^{0\prime} \Gamma_m^{0})^{-1} (f_t^0 H_{it}'{\Sigma}_{\epsilon H_t}^{-1} \Gamma_m^{0} + \Gamma_m^{0\prime} {\Sigma}_{\epsilon H_t}^{-1} H_{it}f_{t}^{0\prime})$, and $u_k$ denoting an unit vector with the $k$-th entry being $1$.
	\end{theorem}

The limiting performance of $\widehat{\theta}_k - \theta_k^0$ is determined by two terms: (i) $D_{k,\ell}Q_{\ell}\big(\widehat{\Gamma}_m - \Gamma_{m}^0(R^{-1}) '\big)u_k $ and (ii) $D_{k,\ell}Q_{\ell}\Gamma_{m}^0 (R^{-1}-I_r)'u_k$, both contributing to the asymptotic covariance. Here, we not that, although the rotation matrix $R$ is asymptotically identical, the limiting distribution of $\widehat{\theta}_k - \theta_k^0$ is affected by $R$ through term (ii) because the convergence rate of $R$ and $\widehat{\theta}_k$ are of the same order $\sqrt{NT}$.

In the QCF model, the index parameter $\theta_k$ summarize the direction and strength of the relationship between observed characteristics and the latent loading function. To quantify the importance of certain characteristics on the loading function, practitioners can perform Wald tests or construct confidence intervals based on the above CLT directly. To this end, we propose an estimator for $\Xi_{\theta,\tau}$. 

Similarly to the existing quantile literature, the conditional density function $g_{it}(0)$ is involved in the asymptotic covariance, which is hard to estimate in practice. Here, instead of estimate $\Sigma_{\epsilon H_t}$ with its sample analogue, we follow \cite{ma2021estimation} and consider the consistent Powell's kernel estimator \citep{RePEc:att:wimass:8818} 
\begin{align*}
    \widehat{\Sigma}_{\epsilon H_t} := \frac{1}{N}\sum_{i=1}^N H_{it}H_{it}' K\bigg(\frac{y_{it} - \widehat{f}_t'\widehat{\Gamma}_m'H_{it}}{h}\bigg),
\end{align*}
where $K(\cdot)$ is the uniform kernel $K(u) := 2^{-1} 1\{|u| \leq 1\}$ and $h$ is a bandwidth. Then, as $\epsilon_{it}$ is assumed to be independent across $i=1,...,N$ and forms a MDS, it is valid to estimate $\Xi_{\theta,\tau}$ by a simple plug-in estimator as
\begin{align*}
       \widehat{\Xi}_{\theta,\tau} :=  \frac{\tau(1-\tau)}{NT} 
            \sum_{i=1}^N\sum_{t=1}^T
             & \widehat{D}_{k,\ell} Q_{\ell}  \Big\{  
             \widehat{\Sigma}_{\epsilon H_t}^{-1} H_{it} \widehat{f}_t^{\prime} 
             + 
             2^{-1} \widehat{\Gamma}_m(\widehat{\zeta}_{it} + \widehat{\zeta}_{it}')
             \Big\} 
             u_k \\
             & \cdot u_k'
              \Big\{  \widehat{f}_t H_{it}'
            \widehat{\Sigma}_{\epsilon H_t}^{-1} 
             + 
             2^{-1} (\widehat{\zeta}_{it} + \widehat{\zeta}_{it}')\widehat{\Gamma}_m^{\prime}
             \Big\}  Q_{\ell}' \widehat{D}_{k,\ell}' ,
\end{align*}
    where $\widehat{D}_{k,\ell}$ is estimated as (\ref{D_hat}), and $\widehat{\zeta}_{it} := (\widehat{\Gamma}_m^{\prime} \widehat{\Gamma}_m)^{-1} (\widehat{f}_t H_{it}'\widehat{\Sigma}_{\epsilon H_t}^{-1} \widehat{\Gamma}_m + \widehat{\Gamma}_m^{\prime} \widehat{\Sigma}_{\epsilon H_t}^{-1} H_{it}\widehat{f}_{t}^{\prime}) $. 

In the following corollary, we establish the consistency of $\widehat{\Xi}_{\theta,\tau}$.
\begin{corollary}\label{t6:Xi_estimates}
    Suppose the assumptions of Theorem~\ref{lemma.A3} hold, and $m^{3/2+9d/4}\ln(T)N^{-1} h ^{-2} = o(1)$, then, as $(N,T) \to (\infty,\infty)$, we have $\widehat{\Xi}_{\theta,\tau} \overset{p}{\to} \Xi_{\theta,\tau}$.
\end{corollary}

Given Corollary~\ref{t6:Xi_estimates}, a standard Wald test can be readily constructed. For instance, researchers may wish to assess whether a subset of $d_1$ ($d_1 < d$) covariates in $x_{it}$ significantly contribute to the loading function $\lambda_k$, by testing the null hypothesis $H_0: S \theta_k^0 = 0$, where $S$ is a $d_1 \times d$ selection matrix with exactly one 1 per row and zeros elsewhere, selecting the $d_1$ components of interest. The Wald statistic is  
$$
    W_{NT} := NT (S \widehat{\theta}_k)'(S \widehat{\Sigma}_{\theta,\tau}S')^{-1} (S \widehat{\theta}_k),
$$
which follows a $\chi^2(d_1)$ distribution asymptotically under $H_0$. When $d_1 = 1$, this reduces to a conventional $t$-type test.


	\section{Simulation Study}\label{sec:simulation}
In this simulation study, we first demonstrate the necessity of incorporating a ridge penalty in the initial step of estimation by comparing the performance of QCF with and without the penalty. We then compare both in-sample and out-of-sample fits of QCF model against the quantile (latent) factor model of \cite{chen2021quantile}, demonstrating the superior performance of QCF when the true factor loadings follow a nonparametric structure.

\subsection{Data Generating Process}
We consider a location-scale model as follows:
	\begin{align*}
		y_{it}  = f_{t(1)} \lambda_{1}(x_{it}'\theta_1) + f_{t(2)} \lambda_{2}(x_{it}'\theta_2) \epsilon_{it},
	\end{align*}
	where $x_{it} = [x_{it(1)},...,x_{it(d)}]'$ with all entries generated independently from $N(0,1)$ and $\epsilon_{it}$ is generated from $N(0,0.5)$. To ensure that the model strictly satisfies the identification conditions (Assumption~\ref{ass.identification}), 
	 we generate the latent factor $F := [f_{(1)}, f_{(2)}]$, where $f_{(k)} := [f_{1(k)},...,f_{T(k)}]'$ for $k=1,2$ as follows. We first generate $\widetilde{f}_{t(1)}$ from $N(0,1)$ and $\widetilde{f}_{t(2)}$ from $|N(0,1)|$, independently for all $t=1,...,T$.  Then, we define $f_{(2)} := T^{1/2}\widetilde{f}_{(2)} /  \|\widetilde{f}_{(2)}\|$, and $f_{(1)} := T^{1/2}\widetilde{f}_{(1)}^{*}/  \|\widetilde{f}_{(1)}^{*}\|$ where $\widetilde{f}_{(1)}^{*} :=   \widetilde{f}_{(1)} -  ( T^{-1} \widetilde{f}_{(1)}' f_{(2)})f_{(2)}$.
	We define $\lambda_1(x) := \sin( x)$ and $\lambda_2(x) := 0.25 + 0.2 \cos( x)$. For the index parameter, we consider an orthonormal pair $\{\theta_1, \theta_2\}$.\footnote{ We generate $\lambda(\cdot)$ and $\theta$ according to Scenario 2 of Appendix \ref{appendix:identification}. It is easy to check that $F$, $\lambda(\cdot)$ and $\theta$ are normalized properly, such that the identification conditions given in Assumption~\ref{ass.identification} are satisfied.}
    Specifically, we let $\theta_1 := \widetilde{\theta}_1/  \|\widetilde{\theta}_1\|$, and $\theta_2 := \widetilde{\theta}_2^{*}/ \|\widetilde{\theta}_2^{*}\|$ with $\widetilde{\theta}^{*}_2 := \widetilde{\theta}_2 -  ( \widetilde{\theta}_2' \theta_1) \theta_1$. In the simulation, we consider two sets of $\{\widetilde{\theta}_1, \widetilde{\theta}_2\}$ as follows:
     \begin{itemize}
        \item Setting 1: $d=10$. Set $\widetilde{\theta}_1 := [3,2,-1,-1,1,0.5,0.5,0.1,0.1,0.1]'$ and $\widetilde{\theta}_2:= [2,2,-1,1,\allowbreak 0.5, -0.1,0.1,0.1,0.05,0.05]'$.
        \item Setting 2: $d=5$. Set $\widetilde{\theta}_1 := [3,2,-1,-1,1]',$ and $\widetilde{\theta}_2:= [2,-1,1,-0.1,0.1]'$.
    \end{itemize}
 
	Given the above data generating process (DGP), since $f_{t(2)}$ and $\lambda_2(x)$ are non-negative,  the conditional $\tau$-quantile of $y_{it}$ has the form
	\begin{align*}
		Q_{y_{it}}\big(\tau | f_{t},x_{it},\lambda(\cdot)\big) =f_{t,\tau}' \lambda_{\tau}(x_{i,t-1}'\theta),
	\end{align*}
	where $f_{t,\tau} = [f_{t(1)}, f_{t(2)}]'$ and $\lambda_{\tau}(x) = [\lambda_{1}(x), Q_{\epsilon}(\tau)\lambda_{2}(x)]'$ when $\tau \neq 0.5$, and  $f_{t,\tau} = f_{t(1)}$, $\lambda_{\tau}(x) = \lambda_{1}(x)$ when $\tau =0.5$. 

\subsection{Error Measurments}
In the simulation study, we consider both in-sample and out-of-sample estimation. Using data $\{(y_{it}, x_{it}),i=1,...,N, t=1,...,T-1\}$, we obtain the in-sample estimation of QCF as 
\begin{align*}
    	\widehat{Q}_{y_{it}}(\tau| \mathcal{F}_{t-1}) = \widehat{f}_{t}' \widehat{\lambda}_{i,t-1},
		\quad 
		i=1,...,N; t=1,...,T-1,
\end{align*}
where $\widehat{\lambda}_{it} := [\widehat{\lambda}_1(x_{it}'\widehat{\theta}_1),...,\widehat{\lambda}_r(x_{it}'\widehat{\theta}_r)]'$.
Then, given $\{\widehat{\lambda}_k(\cdot)\}$ and $\{\widehat{\theta}_k\}$, we obtain the out-of-sample fit of QCF model at time $T$ as
	\begin{align*}
		\widehat{Q}_{y_{iT}}(\tau| \mathcal{F}_{T-1}) =  \widehat{f}_{T}' \widehat{\lambda}_{i,T-1},
        \quad i=1,...,N,
	\end{align*}
where the factor realization $\widehat{f}_{T}$ are constructed from a cross-sectional $\tau$th-quantile regression \citep{koenker1978regression} of $\{y_{iT}\}$ on $\{\widehat{\lambda}_{i,T-1}\}$. We note that $\widehat{f}_T$ is not a forecast but a post-hoc realization conditional on $\widehat{\lambda}_{i,T-1}$ estimated using information up to time $T-1$. This is a common practice in asset pricing literature, not meant to forecast future factor values, but to evaluate how well current risk exposure explains future returns.
For the QFM model, we replace $\widehat{\lambda}_{it}$ with $\widehat{\lambda}_i$ obtained by the iterative quantile regression algorithm of \cite{chen2021quantile}.

	We consider two error measurements in this simulation study. The first is the quantile hitting error (QHE), 
	\begin{align*}
		& \text{QHE}({\tau}) :=  \Big | \frac{1}{NT}\sum_{i=1}^{N}\sum_{t=1}^{T} 1\big\{y_{it} < \widehat{Q}_{y_{it}}(\tau | \mathcal{F}_{t-1})\big\}  - \tau \Big|,
	\end{align*}
	where, with slight abuse of notation, $T$ denotes the total number of testing periods.
	QHE reports the absolute difference between the quantile hitting probability and the theoretical quantile level $\tau$. Ideally, for a $\tau$-quantile, we expect the fitted value to be greater than the response $1-\tau$ fraction of the time. So QHE is a validity check of quantile estimation, often considered in risk management.
	The second is the average quantile empirical error (AQE)
	\begin{align*}
		\text{AQE}({\tau}) := \frac{1}{NT}\sum_{i=1}^{N}\sum_{t=1}^{T}\rho_\tau \big(y_{it} - \widehat{Q}_{y_{it}}(\tau | \mathcal{F}_{t-1})\big),
	\end{align*}
	which quantifies the estimation accuracy under the quantile loss.

	\subsection{QCF Estimation with and without Ridge Penalty}\label{sec:simulation.1}	
	
We compare the in-sample and out-of-sample fits of QCF model with and without the penalty using Setting 1 of the DGP, where the number of characteristics is large but some of the characteristics are weak in explaining the loading dynamics. To ease the computational burden, we perform the variable selection in the first 50 repetitions, and in the remaining repetitions, we set the hyperparameters as the average of those in the first 50 simulations.

Table \ref{table:simulation1} reports the estimation accuracy of the QCF estimation with and without a ridge penalty. Both estimations are consistent as suggested by the asymptotic analysis. Results evaluated by AQE and QHE suggest that while the ridge penalty marginally increases the in-sample AQE in some cases, it consistently improves out-of-sample predictive performance across all quantiles. These findings illustrate the classic bias-variance trade-off, where introducing regularization slightly increases bias but significantly reduces variance, leading to better out-of-sample accuracy.

    	\begin{table}[!htbp]
		\centering
		\caption{QFM Estimation Accuracy With and Without Ridge Penalty}\label{table:simulation1}
		\tabcolsep 0.2in
		\resizebox{0.75\textwidth}{!}{%
			\begin{tabular}{@{} l  c c  | r r  |r  r   @{}} 
				\toprule
				$\tau$& N & T  & 
				\multicolumn{2}{c|}{In-sample Fit} & \multicolumn{2}{c}{Out-of-sample Fit} \\
				&      &         & $a_{NT}=0$ & $a_{NT}= \widehat{a}_{opt}$ & $a_{NT}=0$ &  $a_{NT}= \widehat{a}_{opt}$\\
				\midrule 
				\multicolumn{7}{c}{\textit{Panel A: AQE}}\\
				\midrule
				$0.05$ & 200 & 50 & 0.0447 &  0.0439 &     0.0216 &  0.0206\\
                &  200 & 100 & 0.0445  &0.0442   &  0.0196 & 0.0187 \\
				&   400 & 100 &   0.0439  &0.0434  &   0.0213 & 0.0217    \\
				   $0.50$  & 200 & 50 & 0.1802 &  0.1474   &   0.0954 &  0.0783\\
                   &  200  & 100 &  0.1812 & 0.1480   &  0.0838  &0.0706\\
				&   400 & 100 & 0.1807 & 0.1442  &   0.0904 & 0.0750  \\
				$0.95$ &    200 & 50 &  0.0441 & 0.0434 &     0.0216 & 0.0215\\
                & 200 & 100 &0.0440 & 0.0436   &  0.0191 & 0.0189 \\
				&   400 & 100 &  0.0429  & 0.0435  &   0.0212 & 0.0216 \\
				\midrule 
				\multicolumn{7}{c}{\textit{Panel B: QHE}}\\
				\midrule
				$0.05$ &    200 & 50 &  0.0009  & 0.0024    &  0.0058 & 0.0057 \\
                &  200 & 100 &  0.0012 & 0.0019   &  0.0063 & 0.0055\\
				&   400 & 100 &0.0008  &0.0006 &    0.0033 & 0.0035\\
				$0.50$ &    200 & 50 & 0.0028  & 0.0005    &  0.0360 &  0.0252 \\
                &  200 & 100 &  0.0018  &0.0003    & 0.0299 & 0.0274  \\
				&   400 & 100 &  0.0012  &0.0003    & 0.0251  &0.0224 \\
				$0.95$ &    200 & 50 &  0.0040  & 0.0025   &   0.0079 & 0.0072 \\ 
                &  200 & 100 &  0.0038 & 0.0020 &   0.0077 & 0.0075 \\
				&   400 & 100 &0.0029  &0.0013   &  0.0053 & 0.0050\\
				\bottomrule
		\end{tabular}}
		\begin{minipage}{0.75\linewidth}
			\footnotesize
			\textit{Notes:} The results are obtained based on 100 simulation repetitions. 
		\end{minipage}
	\end{table}

	\subsection{QCF and QFM Model Comparison}
	In this section, we consider Setting 2 in the DGP and compare the performance of QCF against QFM. 
The number of latent factors in QFM is selected using the rank criteria proposed by \cite{chen2021quantile}. Again, we perform the variable selection in the first 50 repetitions, and set the hyperparameters as the average of those in the first 50 simulations in the remaining simulation repetitions.  Tables \ref{table:simulaiton2_AQE} and \ref{table:simulaiton2_QHE} report the in-sample and out-of-sample accuracy of the QCF model against QFM under AQE and QHE measurements. Across all quantiles and sample sizes, the QCF model achieves substantially lower in-sample and out-of-sample AQE and QHE relative to QFM. The improvement is especially pronounced at the tails of the distribution ($\tau=0.05$ and $0.95$), where estimation is generally more challenging even when the sample size is large (for example $N=800$ and $T=100$). For instance, in the case of $\tau = 0.05$ and $N = 400$, the out-of-sample AQE for QCF is approximately $0.0178$ compared to $0.2579$ for QFM. Similarly, out-of-sample QHE for QCF remains below 0.01 across quantiles, whereas QFM exhibits considerably larger errors, particularly at the extremes.

In summary, these results underscore the advantages of modeling factor loadings as functions of observable characteristics. By allowing for semiparametric loading functions, the QCF model captures individual-level variation more effectively than traditional factor models. Furthermore, the benefits of QCF are more pronounced at the tails. Overall, the simulation evidence provides strong support for the superior estimation capabilities of the QCF approach in contexts characterized by nonlinearity and distributional asymmetries.

	\begin{table}[!htbp]
		\centering
		\caption{In-sample and Out-of-sample Point Accuracy Evaluated by AQE}\label{table:simulaiton2_AQE}
		\tabcolsep 0.2in
		\resizebox{0.75\textwidth}{!}{%
			\begin{tabular}{@{}  r r  | c  c |  c c @{}} 
				\toprule
				N & T  & 
				\multicolumn{2}{c|}{In-sample Fit } & \multicolumn{2}{c}{Out-of-sample Fit} \\
				&               & QCF & QFM & QCF & QFM \\
				\midrule 
				$\tau=0.05$ & & & & &   \\
                    200 & 50  &   0.0267  & 0.1933  &0.0178  & 0.2345\\
				200 & 100 &     0.0264 &   0.2127 & 0.0168&  0.2467     \\
				400 & 100 &   0.0257 & 0.2172 & 0.0178 & 0.2579        \\
				800 & 100 &     0.0245 & 0.2188 & 0.0185 & 0.2689  \\
				\midrule
				$\tau=0.50$ & & & & &   \\
				200 & 50 &  0.0778 & 0.1957 & 0.0631 & 0.2367 \\
				200 & 100 &  0.0775  & 0.2152 &  0.0608 &  0.2470   \\
				400 & 100 &   0.0740 & 0.2187 & 0.0628 & 0.2578   \\
				800 & 100 &    0.0744 & 0.2198 & 0.0650 &  0.2693   \\
				\midrule
				$\tau=0.95$ & & & & &   \\
				200 & 50  &   0.0253  & 0.1932  &0.0176 & 0.2334\\
				200 & 100 &    0.0251   & 0.2128 &  0.0167 &  0.2426    \\
				400 & 100 &   0.0243 & 0.2172 & 0.0176 & 0.2549  \\
				800 & 100 &     0.0231 & 0.2189 & 0.0182 &  0.2677    \\
				\bottomrule
		\end{tabular}}
		\begin{minipage}{0.75\linewidth}
			\footnotesize
			\textit{Notes:} The results are obtained based on 250 simulation repetitions.
		\end{minipage}
	\end{table}
	
	\begin{table}[!htbp]
		\centering
		\caption{In-sample and Out-of-sample Point Accuracy Evaluated by QHE}\label{table:simulaiton2_QHE}
		\tabcolsep 0.2in
		\resizebox{0.75\textwidth}{!}{%
			\begin{tabular}{@{}  r r  | c  c |  c c @{}} 
				\toprule
				N & T  & 
				\multicolumn{2}{c|}{In-sample Fit } & \multicolumn{2}{c}{Out-of-sample Fit} \\
				&               & QCF & QFM & QCF & QFM \\
				\midrule 
				$\tau=0.05$ & & & & &   \\
				200 & 50 &   0.0139 & 0.4386 & 0.0053 &  0.4418\\
				200 & 100 &    0.0153  & 0.4421  & 0.0060 &  0.4434    \\
				400 & 100 &    0.0132 & 0.4448 & 0.0034 & 0.4469     \\
				800 & 100 &     0.0116 & 0.4462 & 0.0018   & 0.4470    \\
				\midrule
				$\tau=0.50$ & & & & &   \\
				200 & 50 &  0.0035 & 0.0048 & 0.0270 & 0.0279\\
				200 & 100 &   0.0028  & 0.0032 &  0.0259  & 0.0277   \\
				400 & 100 &       0.0018  & 0.0024  & 0.0191 & 0.0206    \\
				800 & 100 &        0.0012 & 0.0021  & 0.0164   & 0.0139   \\
				\midrule
				$\tau=0.95$ & & & & &   \\
				200 & 50  &  0.0126 & 0.4448  & 0.0070 &  0.4455\\
				200 & 100 &   0.0119 &  0.4445 &  0.0072  & 0.4470   \\
				400 & 100 &    0.0099 & 0.4472 & 0.0042 & 0.4474   \\
				800 & 100 &     0.0088 & 0.4491 & 0.0023&  0.4489    \\
				\bottomrule
		\end{tabular}}
		\begin{minipage}{0.75\linewidth}
			\footnotesize
			\textit{Notes:} The results are obtained based on 250 simulation repetitions. 
		\end{minipage}
	\end{table}

	\section{Understanding the Return Distribution of Corporate Bonds}\label{sec:empirical}
In this section, we analyze the dynamic distributional structure of the U.S. corporate bond returns from FINRA’s	Trade Reporting and Compliance Engine (TRACE) data, publicly available via Wharton Research Data Research Services (WRDS). 
The U.S. corporate bond market is one of the largest and most vital components of global financial markets, with an outstanding value exceeding \$10 trillion as of recent estimates. It plays a crucial role in corporate financing and portfolio diversification. However, the market is also known for its significant tail risk, especially during periods of financial stress, where credit spreads widen abruptly and liquidity deteriorates sharply\citep[e.g.,][]{kargar2021corporate}.

To assess the performance of our proposed QCF model, we conduct both in-sample and out-of-sample evaluations against two benchmarks: (i) the quantile latent factor model (QFM) of \citet{chen2021quantile}, and (ii) quantile regression using observed Fama-French five factors (QR-FF5). In addition, we interpret the latent factors recovered by the QCF model and analyze how observed bond- and firm-level characteristics contribute to the heterogeneity in factor exposures (i.e., factor loadings), thereby offering new insights into the systematic structure of return dynamics across the conditional distribution.

\subsection{Data}
We consider the corporate bonds with a duration no less than 3 years from the WRDS/TRACE dataset.  The return data are collected at a monthly frequency from July 2003 to August 2020, with $N=5346$, and $T=206$. The dataset is processed following the same procedure as \cite{kelly2022reconciling} and \cite{kelly2023modeling}. Specifically, the bond returns are scaled by the previous month's ``Duration times Spread'' (DtS) to have similar ex ante volatility. 
	
There are numerous ``anomaly" characteristics identified in bond return literature. In this paper, we consider the 14 (lagged) firm/bond characteristics that are found to be significant in characterizing the linear time-varying loading functions in \cite{kelly2023modeling}. However, their usefulness in the quantile factor model is thus far unknown. The characteristics can be classified into four categories: bond-specific characteristics, firm-level fundamentals, market-based risk measures, and momentum signals. The detailed descriptions are given in Table~\ref{table:characteristics}. All characteristics are standardized to have a zero mean and a unit variance. 

It is widely supported by empirical evidence that bond returns exhibit heavy-tailed behavior \citep{bessembinder2009measuring}. In our sample, nearly half of the bonds display kurtosis exceeding 5, further suggesting the potential presence of heavy tails. This motivates the consideration of a quantile-based factor model, which not only captures the entire distribution of returns but also offers greater robustness to heavy tails and outliers compared to traditional mean-based factor models.
	
	\begin{table}
		\centering
		\caption{Description of Firm/Bond Characteristics\label{table:characteristics}}
		\tabcolsep 0.2in
		\resizebox{\textwidth}{!}{%
			\begin{tabular}{@{}ll@{}}
\toprule
\multicolumn{2}{c}{\textbf{Bond-Specific Characteristics}} \\
\midrule
Bond age & The term to maturity of a bond, measured in years. \\
Bond volatility & The bond return volatility over the past 24 months. \\
Coupon & The current applicable annual interest rate that the bond’s issuer is obligated to pay the bondholders. \\
Duration & The derivative of the bond value to the credit spread divided by the bond value, calculated by the data vendor's proprietary pricing model. \\
Face value & The face value of bonds. \\
Spread & The option-adjusted spread of the bond. \\
Spread-to-D2D & The option-adjusted spread, divided by one minus the cumulative distribution function of distance-to-default defined by \cite{shumway2001forecasting}. \\
\midrule
\multicolumn{2}{c}{\textbf{Firm-Level Fundamentals}} \\
\midrule
Debt-to-EBITDA & The total debt divided by EBITDA. \\
Equity market cap. & The total market value of the firm's outstanding shares of stock. \\
Firm total debt & The sum of the firm's short-term debt and long-term debt. \\
\midrule
\multicolumn{2}{c}{\textbf{Market-Based Risk Measures}} \\
\midrule
Equity volatility & The stock return volatility over the past 24 months. \\
VIX beta & The sum of coefficients on current and lagged VIX in a regression of bond returns on Mkt-RF, HML, SMB, DEF, TERM, VIX, and lagged VIX, run over the past 60 years. \\
\midrule
\multicolumn{2}{c}{\textbf{Momentum Signals}} \\
\midrule
Mom. 6m equity & The 6-2 momentum of the firm's stock returns. \\
Mom. 6m industry & The industry-adjusted 6-2 momentum of the bond returns. \\
\bottomrule
\end{tabular}
}
\end{table}

\subsection{In-Sample and Out-of-Sample Model Comparison}
  
We first compare the estimation accuracy of our QFM specification to two benchmark models (QFM and QR-FF5) in the literature. 
The out-of-sample fit is performed on a rolling-window basis. Following the same spirit as \cite{kelly2023modeling}, we estimate the factor loading function and index parameters using the most recent 36 months of data, and then estimate the factor realization via simple quantile regression. That is, the one-month-ahead estimation of QCF is given as
	\begin{align*}
		\widehat{Q}_{y_{i,t+1}}^{QCF}(\tau| \mathcal{F}_{t}) = \widehat{f}_{t+1}' \widehat{\lambda}_{it},
		\quad 
		\widehat{\lambda}_{it} := [\widehat{\lambda}_{1}(x_{it}'\widehat{\theta}_1),...,\widehat{\lambda}_{r}(x_{it}'\widehat{\theta}_r)]'
	\end{align*}
	where $\{\widehat{\lambda}_{it}\}$ are estimated using the data $\{(y_{is}, x_{i,s-1})\}$ of the most recent 36 months up to time $t$ and then the factor realization $\widehat{f}_{t+1}$ are constructed from a simple $\tau$th-quantile regression \citep{koenker1978regression} of $\{y_{i,t+1}\}$ on $\{\widehat{\lambda}_{it}\}$. For the QFM model, one-month-ahead estimation is 
	\begin{align*}
		\widehat{Q}_{y_{i,t+1}}^{QFM}(\tau| \mathcal{F}_{t}) = \widehat{f}_{t+1}' \widehat{\lambda}_{i}, 
	\end{align*}
	where $\{\widehat{\lambda}_{i}\}$  are estimated by \cite{chen2021quantile} using a 36-month tailing-window up to time $t$ and then the factor realization $\widehat{f}_{t+1}$ are constructed from quantile regression of $\{y_{i,t+1}\}$ on $\{\widehat{\lambda}_{i}\}$. For the observable factor model QR-FF5, the model is estimated with quantile regression by replacing the latent factors in QFM with the Fama-French 5 factors.
	The first out-of-sample test observation is 36 months after the start of our sample. The out-of-sample testing periods are from July 2006 to August 2020.
	
	To compare the estimation accuracy, we report five error measures.  In addition to QHE and AQE considered in the simulation study, we consider three 
    measurements based on the quantile goodness-of-fit measure ($R^1$) of \cite{koenker1999goodness}. These three metrics can be viewed as quantile analogies of the total $R^2$, time-series $R^2$, and cross-sectional $R^2$ measures in the  cross-sectional asset pricing literature \citep{kelly2023modeling}
	The first $R^1$ measure is ``Total $R^1$'', which aggregates over all bonds and testing periods
	\begin{align*}
		\text{Total } R^1(\tau) & := 1 - \frac{\sum_{i=1}^{N}\sum_{t=1}^{T}\rho_{\tau}\big(y_{it} -  \widehat{Q}_{y_{it}}(\tau | \mathcal{F}_{t-1}))\big)}
		{\sum_{i=1}^{N}\sum_{t=1}^{T}\rho_{\tau}\big(y_{it} \big)},
	\end{align*}
    where $N$ is the total number of bonds under test and $T$ is the total number of testing periods.
    
	The second one is ``Time Series $R^1$''.  Since the underlying bonds vary over time, we define the time series $R^1$ for bond $i$ as 
	\begin{align*}
		R^1_i(\tau) & := 1 - \frac{\sum_{t=1}^{T_i}\rho_{\tau}\big(y_{it} -  \widehat{Q}_{y_{it}}(\tau | \mathcal{F}_{t-1}))\big)}
		{\sum_{t=1}^{T_i}\rho_{\tau}\big(y_{it} \big)},
	\end{align*}
	where $T_i$ is the total number of testing periods for bond $i$, and report the weighted-average time-series $R^1$:
	\begin{align*}
		\text{Time Series } R^1(\tau) & := \frac{1}{\sum_{i=1}^{N}}\sum_{i=1}^{N} T_i R_{i}^1(\tau).
	\end{align*}
	The last error metric is the ``Cross Section $R^1$'', a quantile analogy of the $R^2$ from Fama-MacBeth cross-sectional regressions of test asset returns at time $t$ on conditional beta's as of time $t-1$,
	\begin{align*}
		R^1_t(\tau) & := 1 - \frac{\sum_{i=1}^{N_t}\rho_{\tau}\big(y_{it} -  \widehat{Q}_{y_{it}}(\tau | \mathcal{F}_{t-1}))\big)}
		{\sum_{i=1}^{N_t}\rho_{\tau}\big(y_{it} \big)},
	\end{align*}
	where $N_t$ is the total number of tested bonds at time $t$, and report the simple average:
	\begin{align*}
		\text{Cross Section } R^1(\tau) & := \frac{1}{T}\sum_{t=1}^{T} R_{t}^1(\tau).
	\end{align*}
For all error measurements, if data is missing for bond $i$ at time $t$, we treat $y_{it} = \widehat{Q}_{y_{it}}(\tau | \mathcal{F}_{t}) =0$ in the calculation.

\begin{table}[!htbp]
		\centering
		\caption{Comparative In-Sample Estimation Performance}\label{empirical:table1}
		\tabcolsep 0.2in
		\resizebox{0.55\textwidth}{!}{%
			\begin{tabular}{@{} l   r r r @{}} 
				\toprule
				$\tau$  & QCF&  QFM & QR-FF5\\
				\midrule 
				\multicolumn{4}{c}{\textit{Panel A: QHE}} \\
				\midrule
				0.05  & 0.0037 &  0.0691  &0.3071\\   
				0.50  &0.0396 & 0.0051 & 0.0765\\
				0.95 & 0.0067 & 0.0209 & 0.4397\\ 
				\midrule
				
				\multicolumn{4}{c}{\textit{Panel B:   AQE}}\\
				\midrule
				0.05  &  0.0035 & 0.0036  &0.0085\\   
				0.50  & 0.0086 & 0.0075 & 0.0104\\
				0.95 &  0.0026 & 0.0028 & 0.0102\\ 
				\midrule
				
				\multicolumn{4}{c}{\textit{Panel C:   Total $R^1$}}\\
				\midrule
				0.05 & 0.6528 & 0.6454  &  0.1615\\
				0.50 &  0.2009 & 0.3173 &  0.0564\\
				0.95 & 0.7732  & 0.7684 &  0.1383  \\
				\midrule
				\multicolumn{4}{c}{\textit{Panel D: Time Series  $R^1$} } \\
				\midrule
				0.05 & 0.6313 & 0.5991    &  0.1646\\
				0.50 &  0.1707 & 0.3071 & 0.0568\\
				0.95 & 0.7652  &0.7645    & 0.1369\\
				\midrule
				\multicolumn{4}{c}{\textit{Panel E: Cross Section   $R^1$} } \\
				\midrule
				0.05 & 0.4629 & 0.4103 & 0.0492\\
				0.50 &  0.1527 & 0.2232 & 0.0473\\
				0.95 & 0.6395 &  0.6246 &  0.1122\\
				\bottomrule
			\end{tabular}
		}
		\begin{minipage}{0.75\linewidth}
			\footnotesize
			\textit{Notes:} Sample period: July 2003 to August 2020. 
		\end{minipage}
	\end{table}

    Table~\ref{empirical:table1} reports in-sample model fits using the entire dataset.
    The QCF model identifies three latent factors at $\tau=0.05$ and four at $\tau=0.5$ and 0.95. Loading functions are quadratic ($m=2$) at the tails ($\tau=0.05, 0.95$) but linear ($m=1$) at the median, indicating greater nonlinearity in the tails compared to the central quantile. Interpretations for the identified factors are provided in the next section.
    
    Among the three quantile regression models under consideration, our proposed QCF model demonstrates superior performance across all evaluation criteria at the tail quantiles ($\tau=0.05$ and $0.95$). It consistently achieves the lowest AQE and QHE, indicating that it not only minimises the quantile loss effectively but also provides well-calibrated and reliable probabilistic estimations across quantiles. 
    Furthermore, the QCF model exhibits the highest $R^1$ values, suggesting that it captures a larger proportion of the conditional variation in returns, analogous to greater explanatory power in traditional mean regression. In comparison, the QFM model -- while also incorporating data-driven latent factors and asset-specific loadings -- assumes linearity across quantiles and generally underperforms QCF at the tails. This highlights the advantage of allowing for nonlinearity and covariate-dependent factor structures, particularly in tail regions where return distributions are more volatile and sparse.

{On the other hand, in the central part of the distribution, typically associated with normal market conditions, the QCF model identifies approximately linear loading functions, suggesting limited nonlinear dependence on characteristics. In such underlying conditions, the simpler quantile factor model (QFM), which assumes constant loadings across characteristics, might perform even better due to less overfitting. Therefore, while the QCF model demonstrates superior performance across the entire distribution, especially in the tails where nonlinearity is more pronounced, the QFM model retains an advantage in modeling bond returns near the median level, where the benefits of modeling nonlinear loadings appear limited.}
 
    Finally, the two latent factor models, QF and QFM, consistently outperform the observable factor model, QR-FF5, across all quantiles, as evidenced by lower AQE and QHE and higher total, time series, and cross-sectional $R^1$ values. While QR-FF5, a direct quantile extension of the Fama-French five-factor model, preserves interpretability and economic intuitions, its rigid structure limits its ability to capture cross-sectional heterogeneity in quantile behaviour, especially in the tails. In contrast, data-driven, quantile-dependent latent factors offer greater flexibility in modeling the full distribution of bond returns. In next subsection \ref{inter}, we give more detailed economic interpretation of the potential nonlinear tail risk behaviors with the link to observed bond-characteristics. 

    Table~\ref{empirical:table2} reports the out-of-sample model fits using a rolling 36-month estimation window. Conclusions remain largely consistent with the findings in the sample. However, the improvement in predictive accuracy for QCF over QFM across all quantiles is even more pronounced out-of-sample, underscoring the value of incorporating observed characteristics to capture heterogeneity and time variation in factor loadings. While QCF and QFM yield relatively stable results across both tables, the $R^1$ values for QR-FF5 turn negative out-of-sample. For observable factor models, the only difference between the two tables is that in-sample fits use risk exposure $\widehat{\lambda}_i$ from the full time series asset-by-asset, while out-of-sample fits use rolling $\widehat{\lambda}_i$. The negative $R^1$ values thus reflect a mismatch between historical and future risk exposures for observed factors, a phenomenon well documented in the empirical asset pricing literature \citep[e.g.,][]{kelly2023modeling}.

	In summary, the results underscore the advantage of incorporating both latent factor structures and covariate-dependent nonlinearities, as in QCF, to achieve superior performance in modelling the tail performance of bond return, which is particularly relevant to periods of market booms and busts.

	\begin{table}[!htbp]
		\centering
		\caption{Comparative Out-of-Sample Estimation Performance}\label{empirical:table2}
		\tabcolsep 0.2in
		\resizebox{0.55\textwidth}{!}{%
			\begin{tabular}{@{} l   r r r @{}} 
				\toprule
				$\tau$ & QCF  &  QFM & QR-FF5\\
				\midrule 
				\multicolumn{4}{c}{\textit{Panel A: QHE}} \\
				\midrule
				0.05 & 0.0001 &  0.1006  &0.3446\\   
				0.50  & 0.0001 & 0.0055   & 0.0780\\
				0.95 &  0.0001 & 0.0280  & 0.4589\\ 
				\midrule
				
				\multicolumn{4}{c}{\textit{Panel B:   AQE}}\\
				\midrule
				0.05 & 0.0026  & 0.0040  &0.0126\\   
				0.50 &  0.0078  & 0.0078 & 0.0125\\
				0.95 & 0.0023 & 0.0028   & 0.0146\\ 
				\midrule
				
				\multicolumn{4}{c}{\textit{Panel C: Total $R^1$}}\\
				\midrule
				0.05 & 0.7454 &  0.6042   &  -0.2241\\
				0.50 &  0.2792& 0.2798    &  -0.1476\\
				0.95 &  0.8026  & 0.7598  &  -0.2550  \\
				\midrule
				\multicolumn{4}{c}{\textit{Panel D: Time Series  $R^1$} } \\
				\midrule
				0.05 &  0.6768 & 0.5067     &  -0.4655 \\
				0.50 & 0.2259& 0.2339   & -0.1880\\
				0.95 &   0.7675  & 0.7448  & -0.3612\\
				\midrule
				\multicolumn{4}{c}{\textit{Panel E: Cross Section   $R^1$} } \\
				\midrule
				0.05 &  0.5102 & 0.3588 &  -0.9596\\
				0.50 &  0.1954& 0.1839 & -0.1932\\
				0.95 &  0.6711  & 0.6243  &  -0.4678\\
				\bottomrule
			\end{tabular}
		}
		\begin{minipage}{0.75\linewidth}
			\footnotesize
			\textit{Notes:} Sample period: July 2003 to August 2020, the first out-of-sample test observation is 36 months after the start of our sample. 
		\end{minipage}
	\end{table}

\subsection{Further Interpretations: Characteristics and Factors}\label{inter}
For practitioners, it is critical to 1) interpret the latent factors and 2) understand how bond characteristics shape the risk exposure in the QCF model. In this section, we answer the above two questions by examining the latent factor and single index estimates. 

Recall that the model’s identification strategy ensures that each index $\theta_k$ is normalized (i.e., has unit norm), and all observed characteristics are standardized to have zero mean and unit variance. This normalization enables a direct interpretation of the relative influence and directional contribution of each characteristic in shaping the latent factor exposures across quantiles in terms of the magnitude of the index. Note that the QCF model's index parameters do not directly reflect marginal return sensitivities. These sensitivities depend on the index parameters, the nonlinear loading functions' values and derivatives, and interactions among all factors and loadings.

\begin{figure}[htbp]
    \centering
\caption{In-sample Factor and Index Parameter Estimates at $\tau=0.05$} \label{figure:ft_theta_q5}
    \caption*{(i) Factor Estimates $\widehat{f}_{t(k)}$}
    
    \begin{minipage}{0.32\textwidth}
        \centering
        \includegraphics[width=\linewidth]{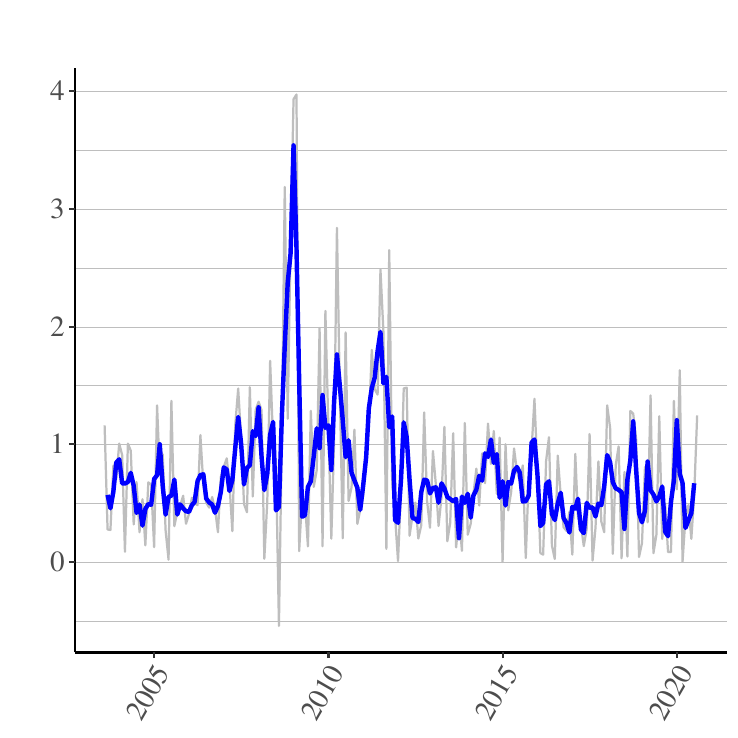}
        \caption*{$k=1$}
    \end{minipage}
    \begin{minipage}{0.32\textwidth}
        \centering
        \includegraphics[width=\linewidth]{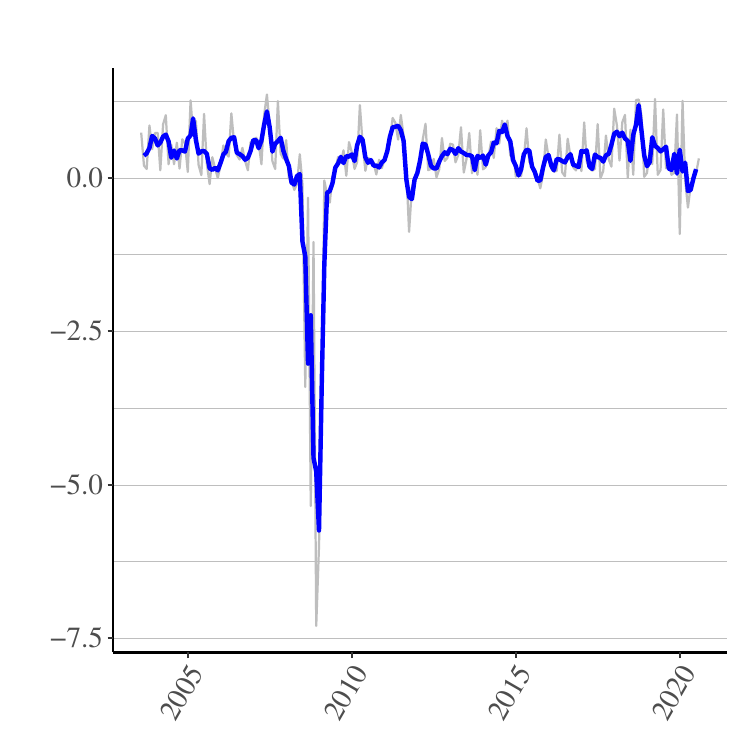}
        \caption*{$k=2$}
    \end{minipage}
    \begin{minipage}{0.32\textwidth}
        \centering
        \includegraphics[width=\linewidth]{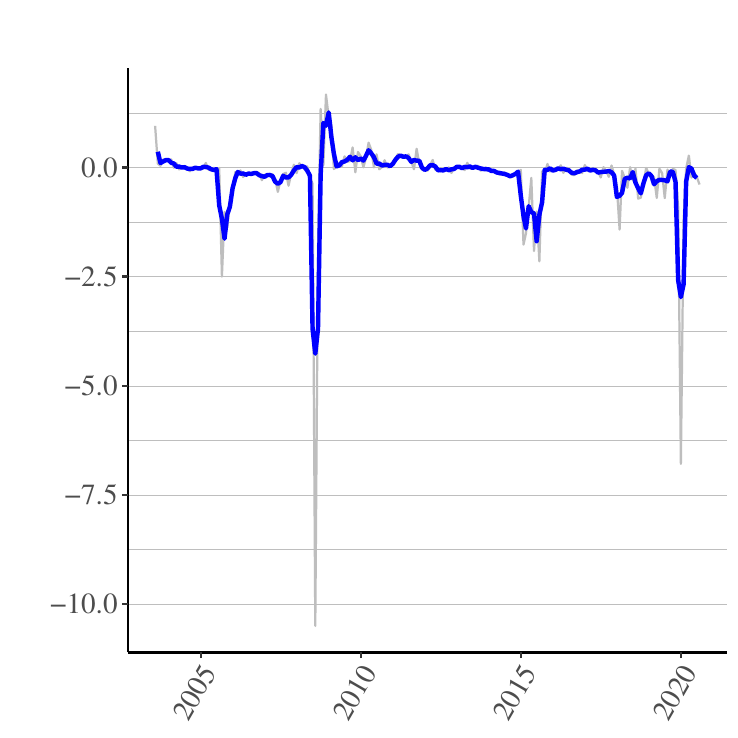}
        \caption*{$k=3$}
    \end{minipage}

    \vspace{1cm}
    \caption*{(ii) Index Parameter Estimates $\widehat{\theta}_k$}

    \begin{minipage}{0.32\textwidth}
        \centering
        \includegraphics[width=\linewidth]{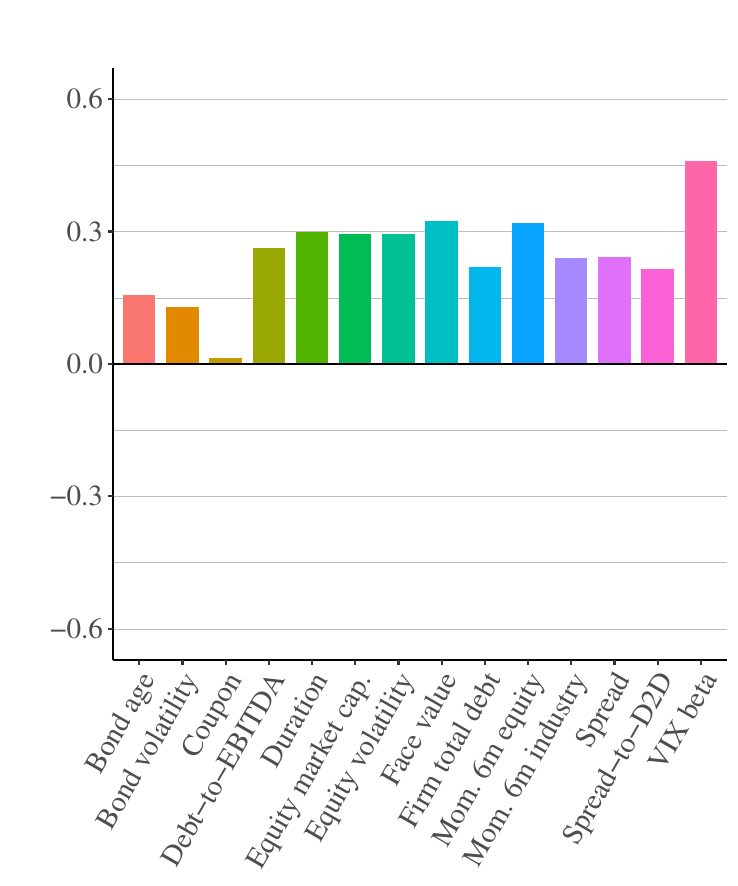}
        \caption*{$k=1$}
    \end{minipage}
    \begin{minipage}{0.32\textwidth}
        \centering
        \includegraphics[width=\linewidth]{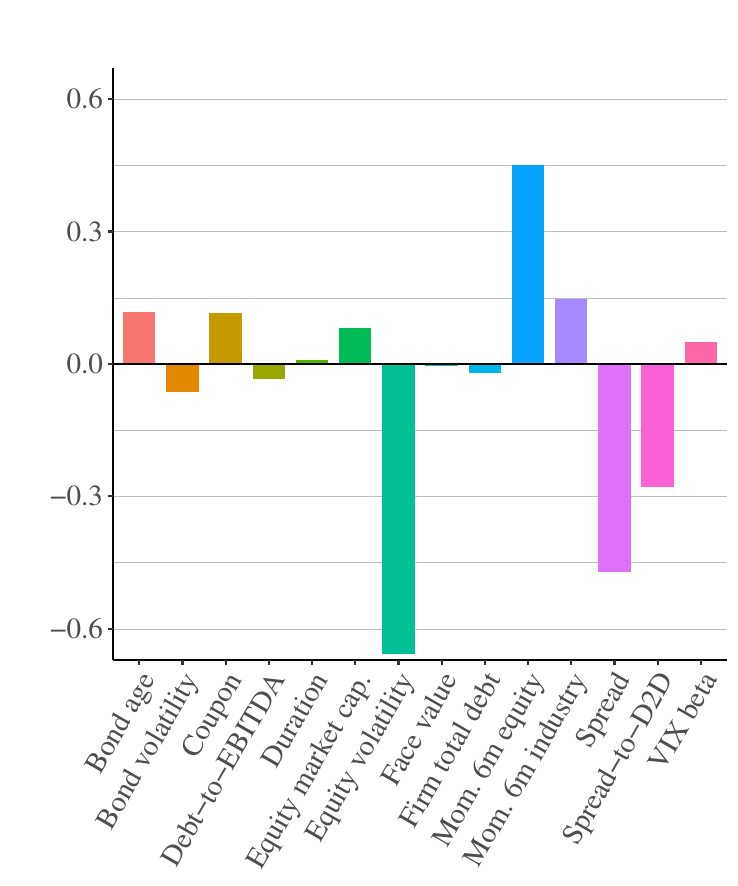}
        \caption*{$k=2$}
    \end{minipage}
    \begin{minipage}{0.32\textwidth}
        \centering
        \includegraphics[width=\linewidth]{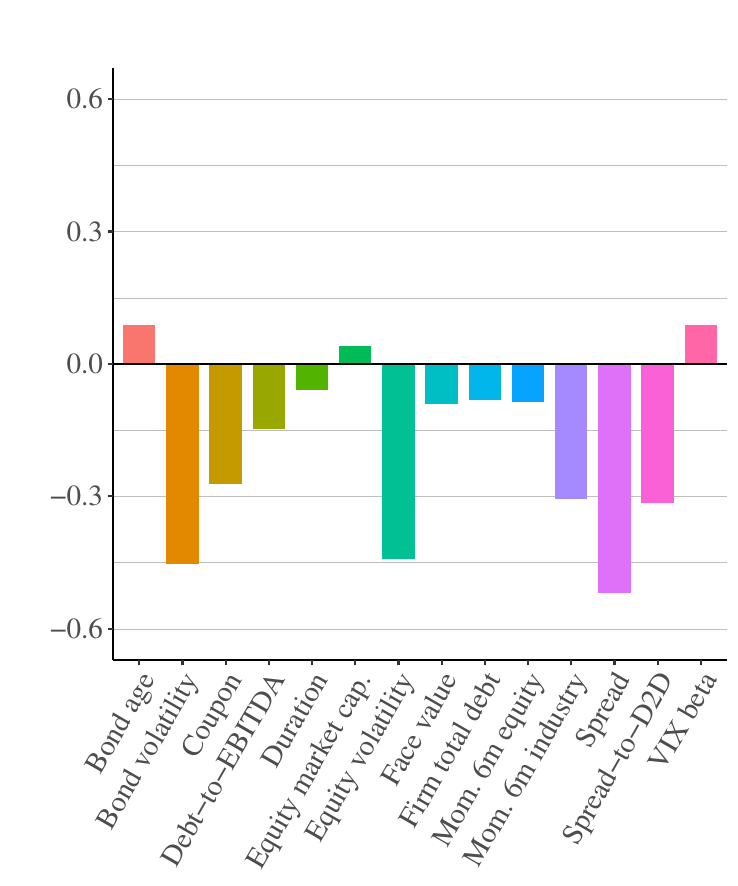}
        \caption*{$k=3$}
    \end{minipage}
    \begin{minipage}{\linewidth}
			\footnotesize
			\textit{Notes:} This figure presents the in-sample estimates for the latent factors and single index parameters at $\tau = 0.05$ using data from July 2003 to August 2020. The top panel reports the factor estimates. The raw estimates are plotted in grey, while the smoothed estimates with Moving Average of order 3 (MA(3)) are plotted in blue. The bottom panel presents the single index estimates. 
		\end{minipage}
\end{figure}

\begin{figure}[htbp]
    \centering
\caption{In-sample Factor and Index Parameter Estimates at $\tau=0.5$} \label{figure:ft_theta_q50}
    \caption*{(i) Factor Estimates $\widehat{f}_{t(k)}$}
    \begin{minipage}{0.24\textwidth}
        \centering
        \includegraphics[width=\linewidth]{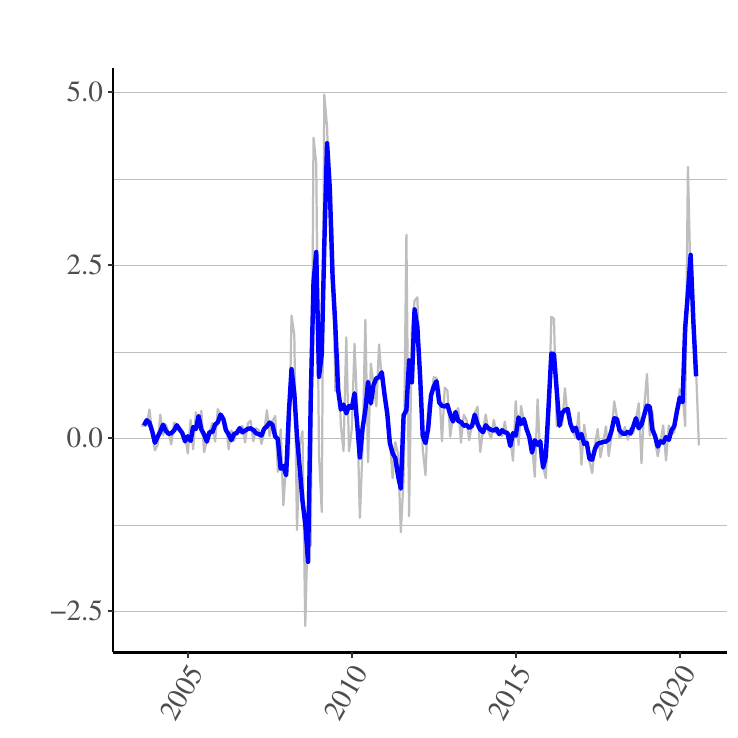}
        \caption*{$k=1$}
    \end{minipage}
    \begin{minipage}{0.24\textwidth}
        \centering
        \includegraphics[width=\linewidth]{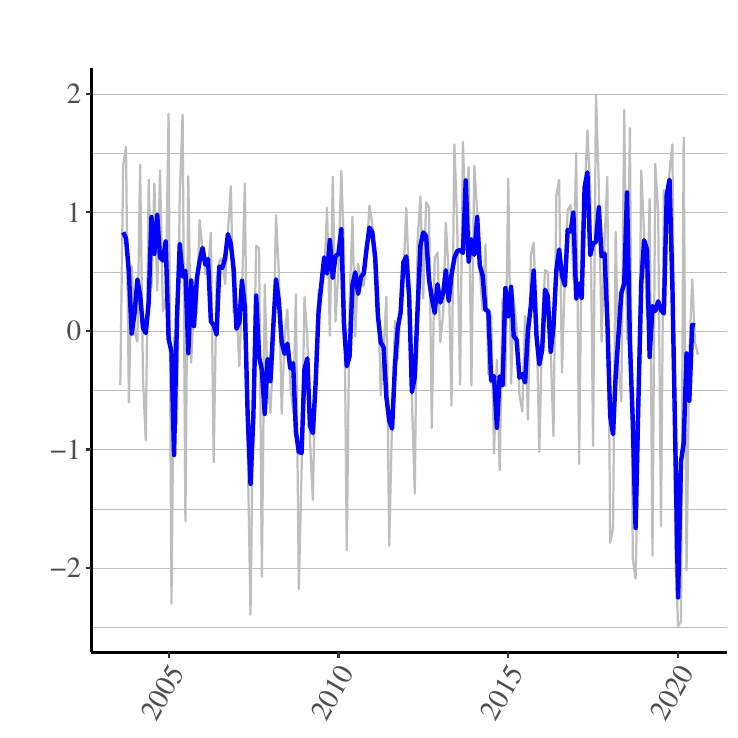}
        \caption*{$k=2$}
    \end{minipage}
    \begin{minipage}{0.24\textwidth}
        \centering
        \includegraphics[width=\linewidth]{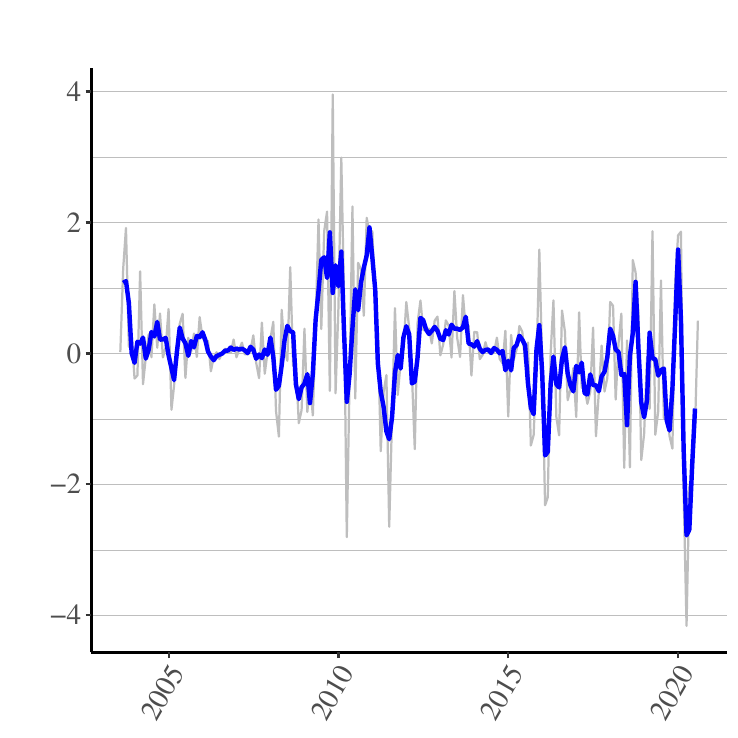}
        \caption*{$k=3$}
    \end{minipage}
\begin{minipage}{0.24\textwidth}
        \centering
        \includegraphics[width=\linewidth]{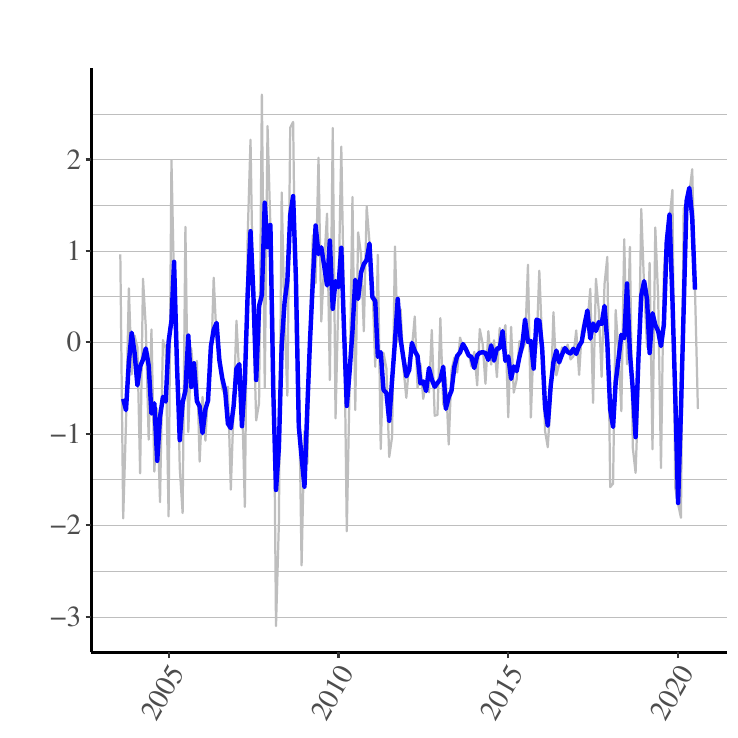}
        \caption*{$k=4$}
    \end{minipage}

    \vspace{1cm}
    \caption*{(ii) Index Parameter Estimates $\widehat{\theta}_{k}$}
    \begin{minipage}{0.24\textwidth}
        \centering
        \includegraphics[width=\linewidth]{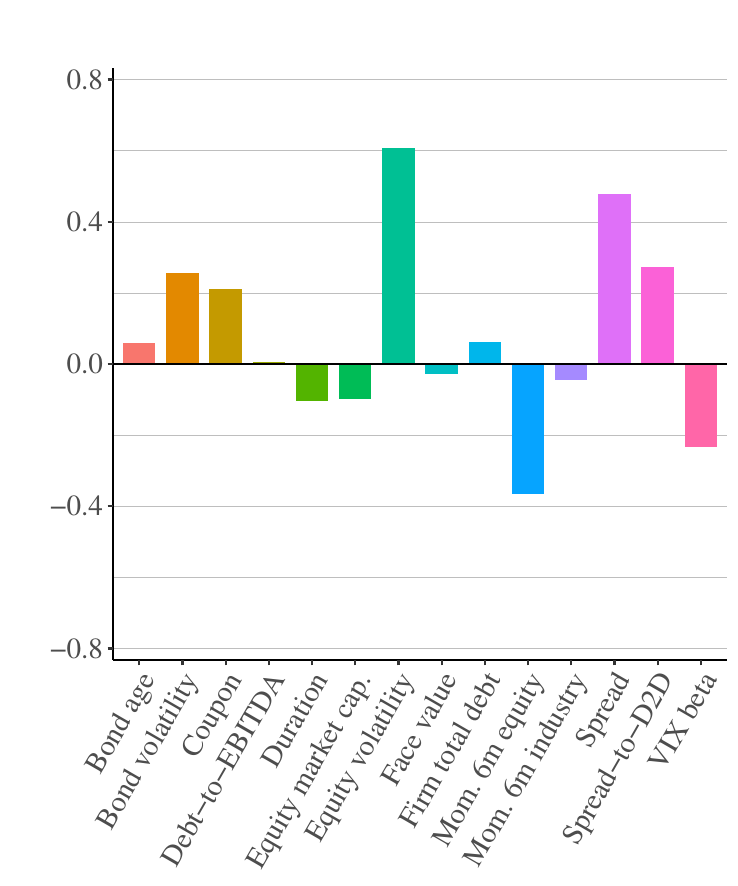}
        \caption*{$k=1$}
    \end{minipage}
    \begin{minipage}{0.24\textwidth}
        \centering
        \includegraphics[width=\linewidth]{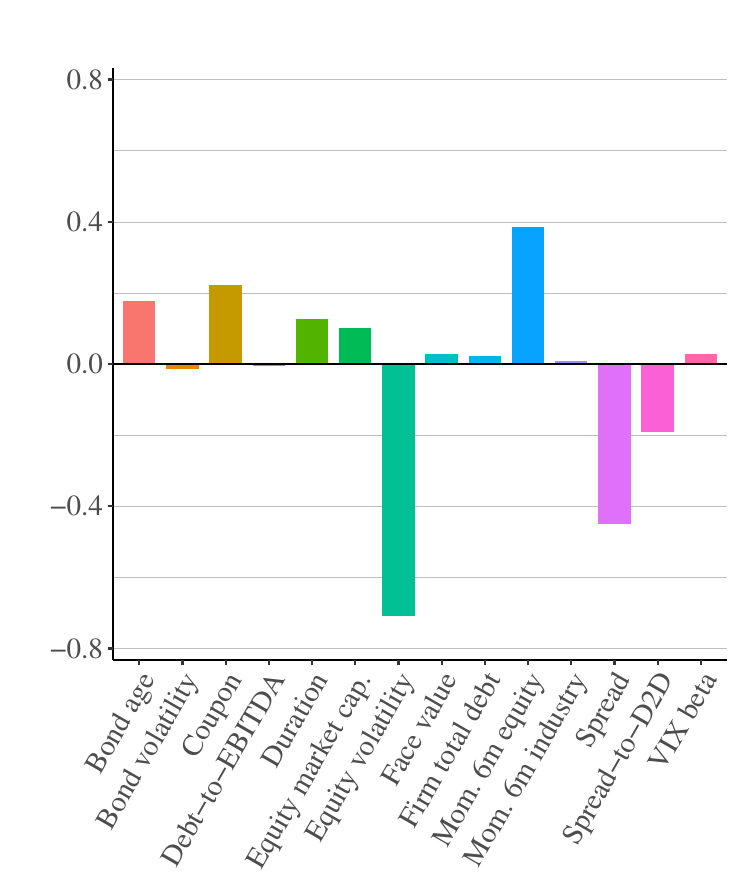}
        \caption*{$k=2$}
    \end{minipage}
    \begin{minipage}{0.24\textwidth}
        \centering
        \includegraphics[width=\linewidth]{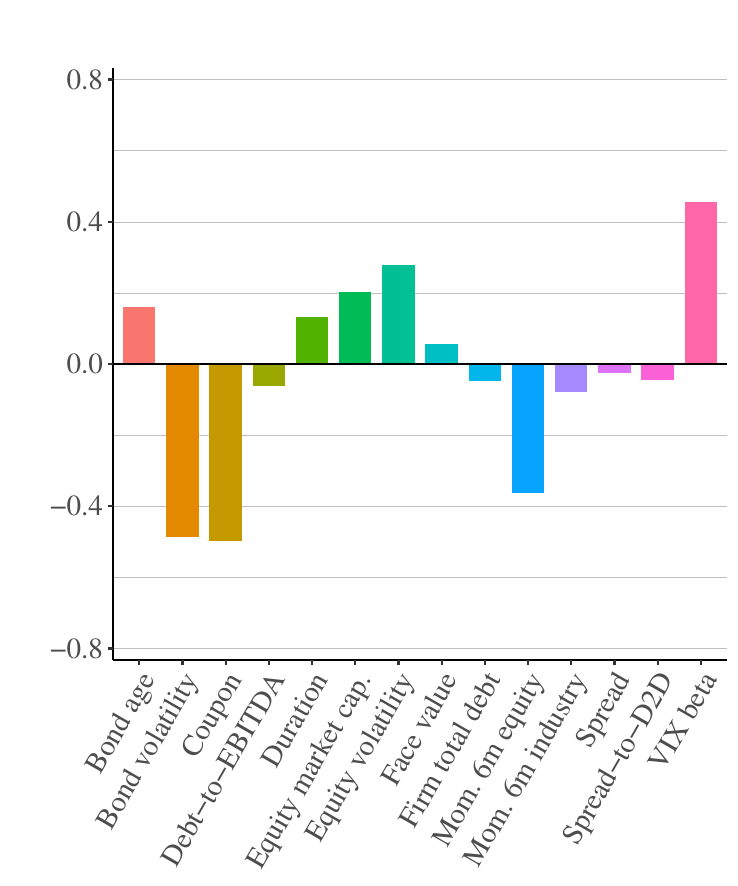}
        \caption*{$k=3$}
    \end{minipage}
    \begin{minipage}{0.24\textwidth}
        \centering
        \includegraphics[width=\linewidth]{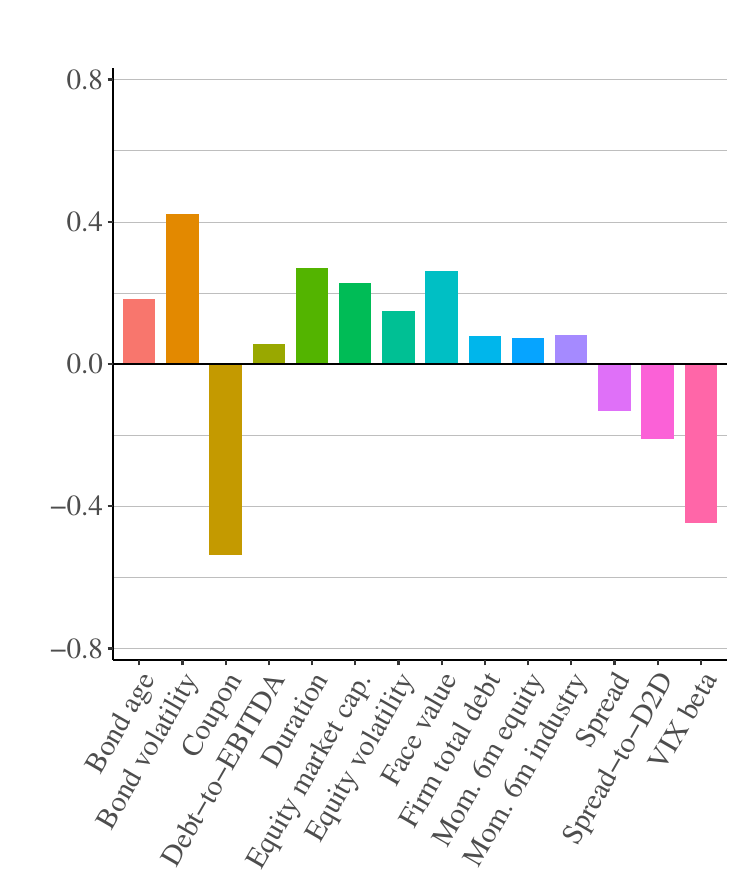}
        \caption*{$k=4$}
    \end{minipage}
    \begin{minipage}{\linewidth}
			\footnotesize
			\textit{Notes:} This figure presents the in-sample estimates for the latent factors and single index parameters at $\tau = 0.5$ using data from July 2003 to August 2020. The top panel reports the factor estimates. The raw estimates are plotted in grey, while the smoothed estimates with MA(3) are plotted in blue. The bottom panel presents the single index estimates. 
		\end{minipage}
\end{figure}

\begin{figure}[htbp]
    \centering
\caption{In-sample Factor and Index Parameter Estimates at $\tau=0.95$} \label{figure:ft_theta_q95}
    \caption*{(i) Factor Estimates $\widehat{f}_{t(k)}$}
    \begin{minipage}{0.24\textwidth}
        \centering
        \includegraphics[width=\linewidth]{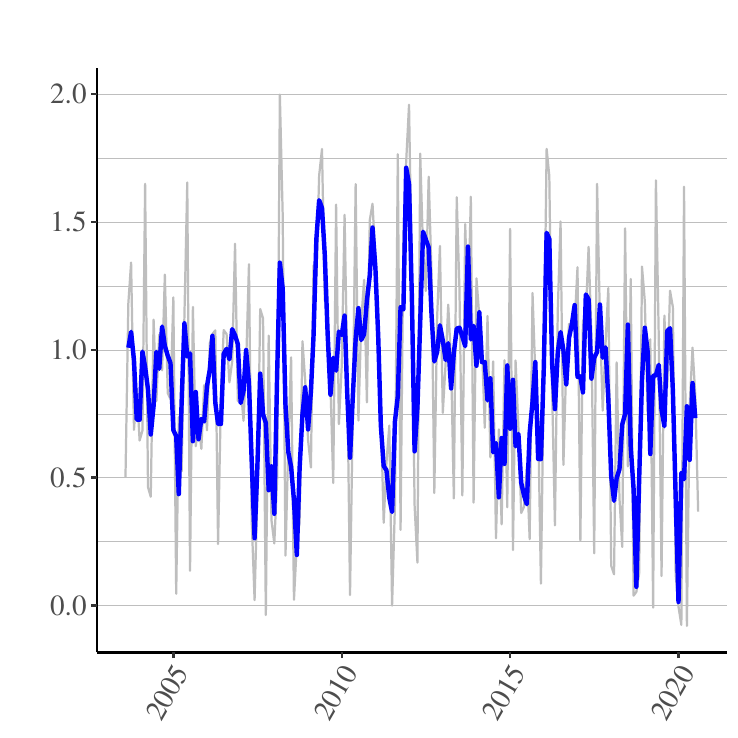}
        \caption*{$k=1$}
    \end{minipage}
    \begin{minipage}{0.24\textwidth}
        \centering
        \includegraphics[width=\linewidth]{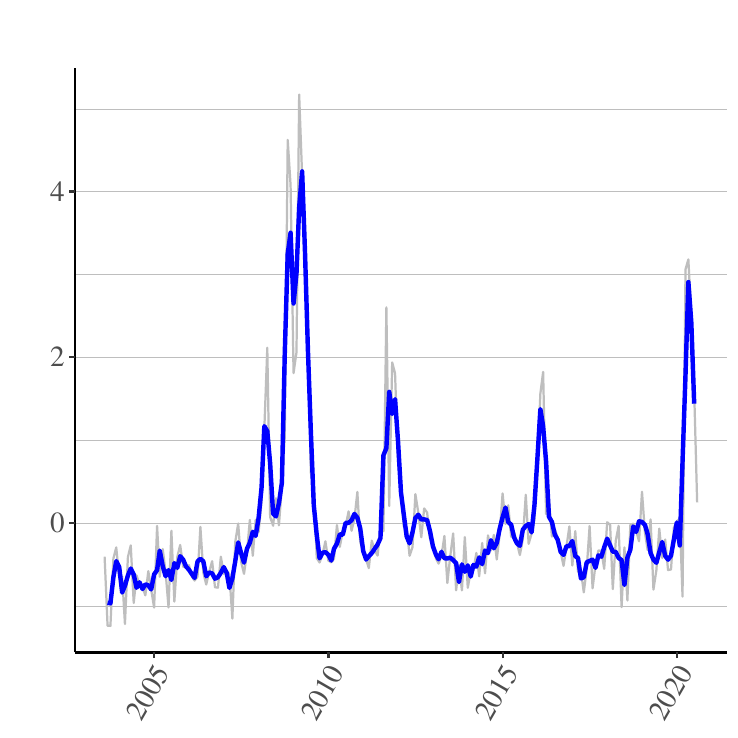}
        \caption*{$k=2$}
    \end{minipage}
    \begin{minipage}{0.24\textwidth}
        \centering
        \includegraphics[width=\linewidth]{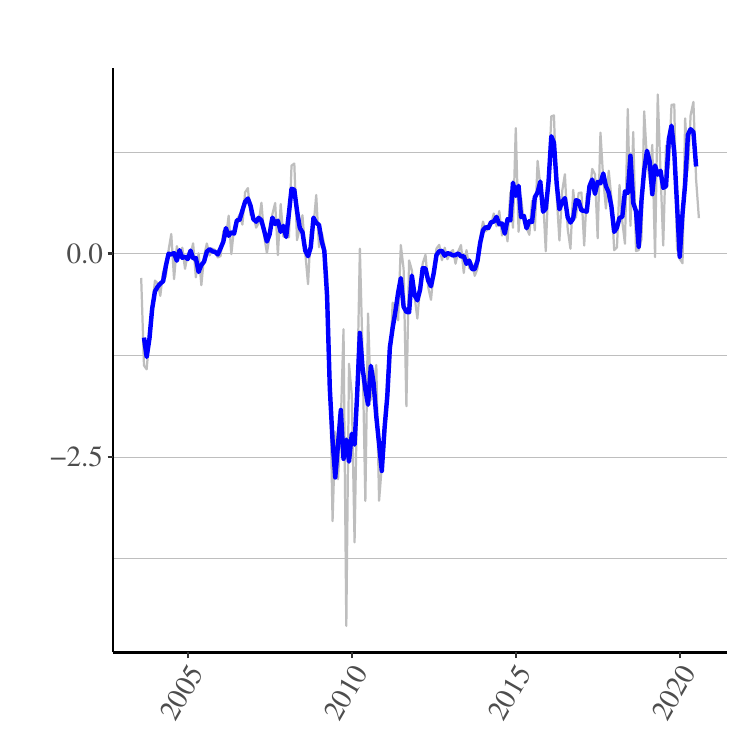}
        \caption*{$k=3$}
    \end{minipage}
\begin{minipage}{0.24\textwidth}
        \centering
        \includegraphics[width=\linewidth]{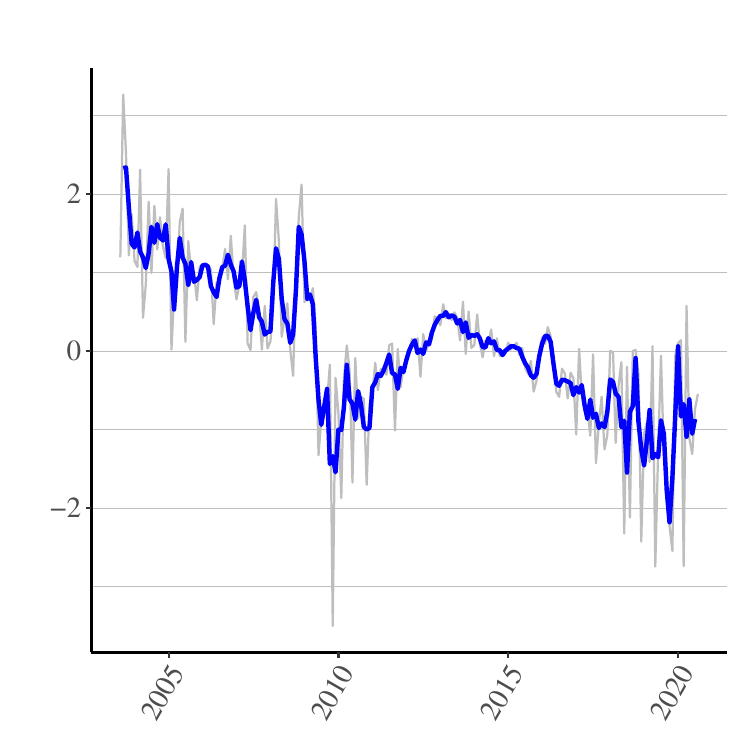}
        \caption*{$k=4$}
    \end{minipage}

    \vspace{1cm}
    \caption*{(ii) Index Parameter Estimates $\widehat{\theta}_{k}$}
    \begin{minipage}{0.24\textwidth}
        \centering
        \includegraphics[width=\linewidth]{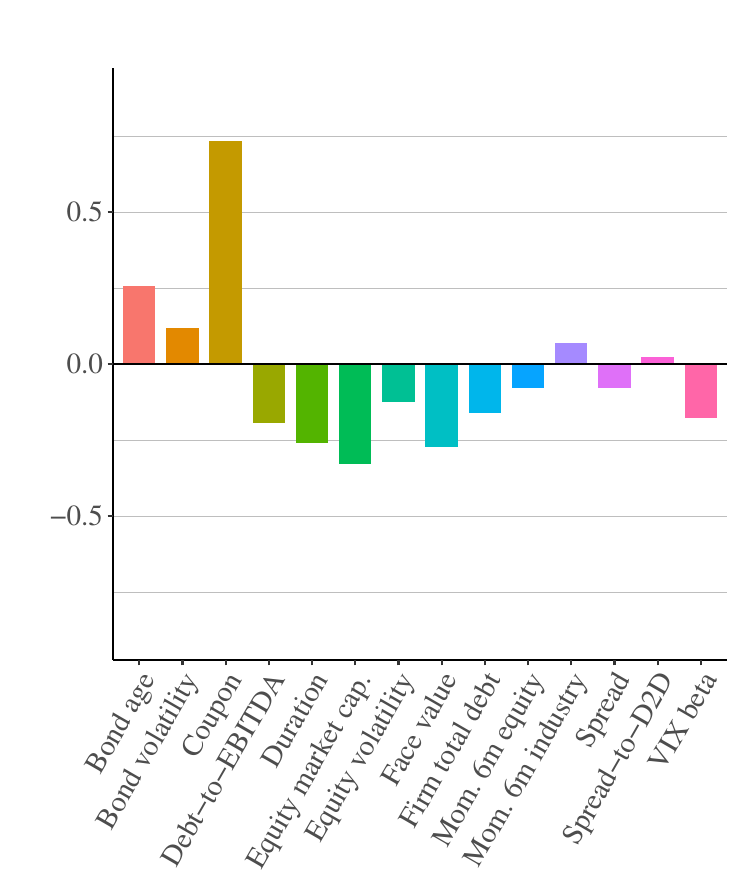}
        \caption*{$k=1$}
    \end{minipage}
    \begin{minipage}{0.24\textwidth}
        \centering
        \includegraphics[width=\linewidth]{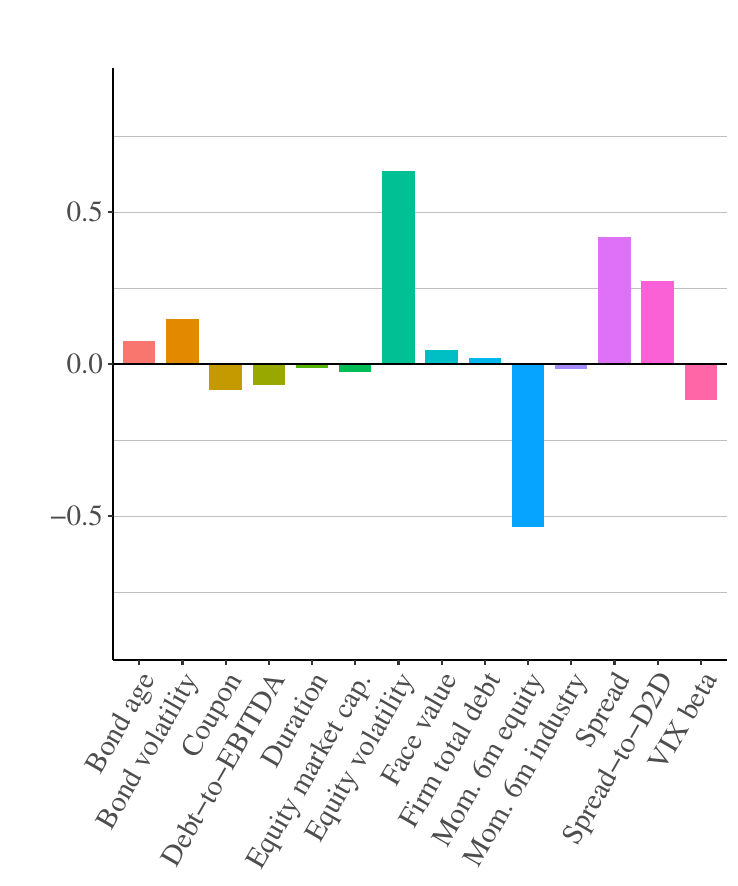}
        \caption*{$k=2$}
    \end{minipage}
    \begin{minipage}{0.24\textwidth}
        \centering
        \includegraphics[width=\linewidth]{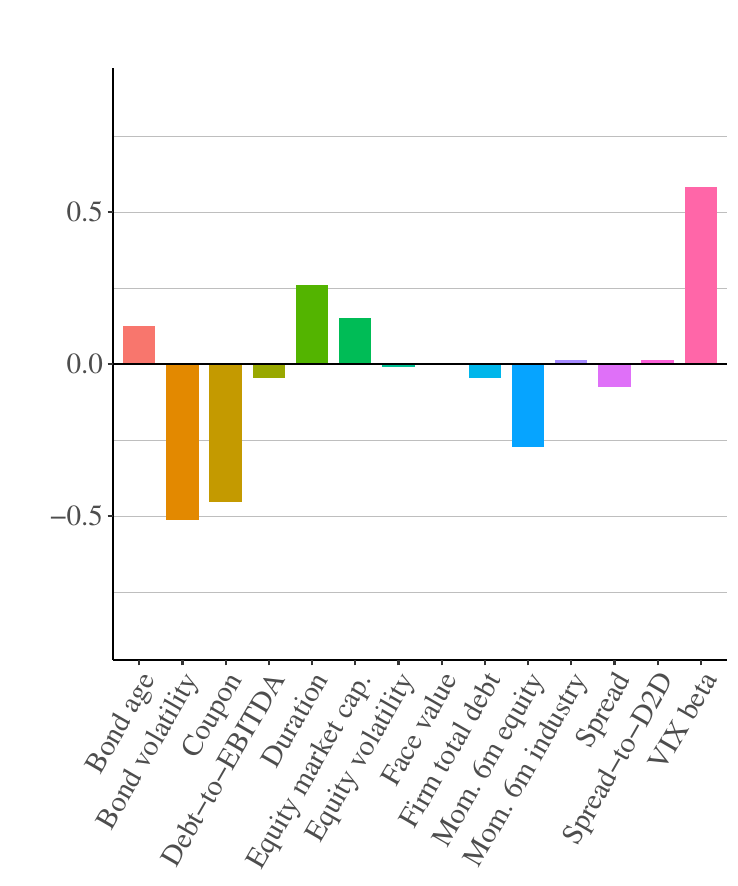}
        \caption*{$k=3$}
    \end{minipage}
    \begin{minipage}{0.24\textwidth}
        \centering
        \includegraphics[width=\linewidth]{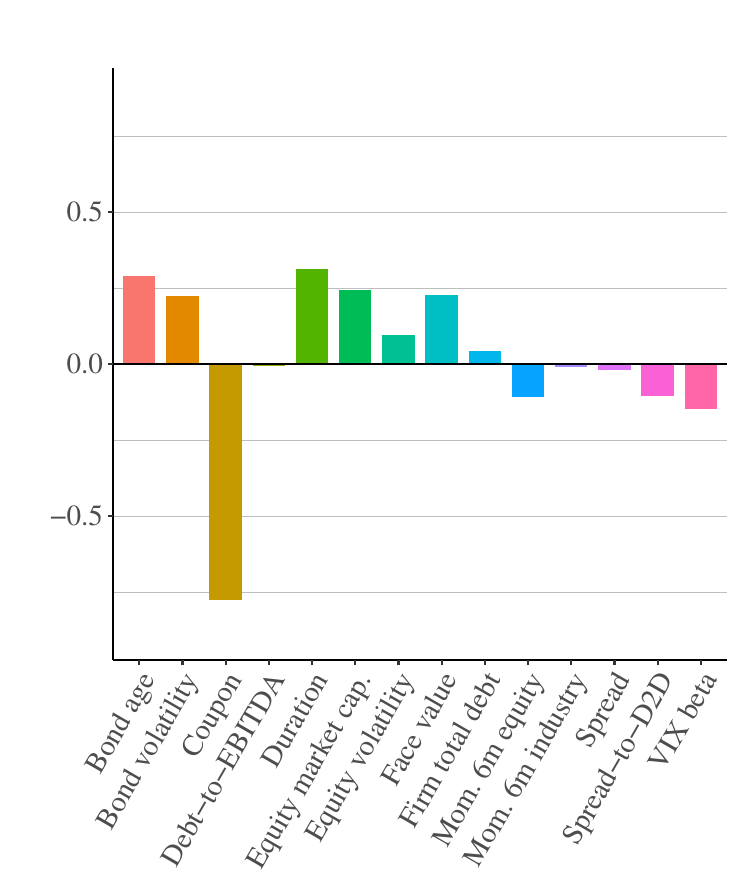}
        \caption*{$k=4$}
    \end{minipage}
    \begin{minipage}{\linewidth}
			\footnotesize
			\textit{Notes:} This figure presents the in-sample estimates for the latent factors and single index parameters at $\tau = 0.95$ using data from July 2003 to August 2020. The top panel reports the factor estimates. The raw estimates are plotted in grey, while the smoothed estimates with MA(3) are plotted in blue. The bottom panel presents the single index estimates. 
		\end{minipage}
\end{figure}

We present the in-sample estimates for factor and index parameters at quantiles $\tau = 0.05,\ 0.5$, and $0.95$ in Figures \ref{figure:ft_theta_q5}-\ref{figure:ft_theta_q95}, respectively. In addition, we formally test the marginal significance of each index parameter, whose null hypothesis is specified as $H_0: \theta^0_{k(j)} = 0$ for a given $k \in \{1,...,r\}$ and $j \in \{1,...,d\}$, using the $t$-statistics constructed under Corollary~\ref{t6:Xi_estimates} and report them in Table~\ref{table:theta}.


    	\begin{table}[!htbp]
		\centering
		\caption{In-sample Index Parameter Estimates}\label{table:theta}
		\tabcolsep 0.2in
		\resizebox{\textwidth}{!}{%
			\begin{tabular}{@{} l | l  l l | l l l l | l l l l  @{}} 
				\toprule
				 &
				\multicolumn{3}{c|}{$\tau=0.05$} & \multicolumn{4}{c|}{$\tau=0.5$} & \multicolumn{4}{c}{$\tau=0.95$}  \\
				Characteristics &       $k=1$ & $k=2$ & $k=3$ & $k=1$ & $k=2$ & $k=3$ & $k=4$   & $k=1$ & $k=2$ & $k=3$ & $k=4$ \\
				\midrule 
  Bond age	&	0.1564			&	0.1181			&	0.0888			&	0.0589			&	0.1772	(	***	)&	0.1615	(	***	)&	0.1826	(	***	)&	0.2554	(	***	)&	0.0758	(	*	)&	0.1240	(	***	)&	0.2900	(	***	)\\		
Bond volatility	&	0.1294			&	-0.0634			&	-0.4532			&	0.2549	(	***	)&	-0.0152			&	-0.4853	(	***	)&	0.4224	(	***	)&	0.1191	(	***	)&	0.1486	(	***	)&	-0.5134	(	***	)&	0.2243	(	***	)\\		
Coupon	&	0.0132			&	0.1155			&	-0.2726			&	0.2101	(	***	)&	0.2209	(	***	)&	-0.4968	(	***	)&	-0.5371	(	***	)&	0.7351	(	***	)&	-0.0856	(	***	)&	-0.4549	(	***	)&	-0.7757	(	***	)\\		
Debt-to-EBITDA	&	0.2628	(	*	)&	-0.0343			&	-0.1466			&	0.0066			&	-0.0058			&	-0.0611			&	0.0565			&	-0.1939	(	***	)&	-0.0690	(	**	)&	-0.0448	(	**	)&	-0.0068			\\		
Duration	&	0.2986	(	**	)&	0.0094			&	-0.0587			&	-0.1048	(	**	)&	0.1269	(	***	)&	0.1321	(	***	)&	0.2704	(	***	)&	-0.2608	(	***	)&	-0.0140			&	0.2614	(	***	)&	0.3122	(	***	)\\		
Equity market cap.	&	0.2939	(	**	)&	0.0816			&	0.0412			&	-0.0992	(	**	)&	0.1017	(	***	)&	0.2013	(	***	)&	0.2284	(	***	)&	-0.3302	(	***	)&	-0.0258			&	0.1522	(	***	)&	0.2424	(	***	)\\		
Equity volatility	&	0.2937	(	*	)&	-0.6579	(	**	)&	-0.4410			&	0.6058	(	***	)&	-0.7096	(	***	)&	0.2781	(	***	)&	0.1500			&	-0.1239	(	***	)&	0.6342	(	***	)&	-0.0097			&	0.0938	(	***	)\\		
Face value	&	0.3239	(	**	)&	-0.0052			&	-0.0908			&	-0.0277			&	0.0279			&	0.0550			&	0.2627	(	***	)&	-0.2736	(	***	)&	0.0463			&	-0.0024			&	0.2263	(	***	)\\		
Firm total debt	&	0.2192	(	*	)&	-0.0194			&	-0.0808			&	0.0628			&	0.0218			&	-0.0492			&	0.0773	(	*	)&	-0.1603	(	***	)&	0.0207			&	-0.0459	(	**	)&	0.0417			\\		
Mom. 6m equity	&	0.3196			&	0.4505			&	-0.0852			&	-0.3662	(	***	)&	0.3843	(	***	)&	-0.3619	(	***	)&	0.0731	(	***	)&	-0.0790	(	***	)&	-0.5382	(	***	)&	-0.2736	(	***	)&	-0.1090	(	***	)\\		
Mom. 6m industry	&	0.2393			&	0.1474			&	-0.3049			&	-0.0461			&	0.0082			&	-0.0785	(	**	)&	0.0817	(	***	)&	0.0685	(	***	)&	-0.0156			&	0.0139			&	-0.0102			\\		
Spread	&	0.2423			&	-0.4717	(	*	)&	-0.5190			&	0.4792	(	***	)&	-0.4509	(	***	)&	-0.0255			&	-0.1332			&	-0.0778	(	***	)&	0.4195	(	***	)&	-0.0750	(	***	)&	-0.0209			\\		
Spread-to-D2D	&	0.2142			&	-0.2793			&	-0.3149			&	0.2726	(	***	)&	-0.1911	(	***	)&	-0.0450			&	-0.2104	(	***	)&	0.0222			&	0.2735	(	***	)&	0.0126			&	-0.1063	(	***	)\\		
VIX beta	&	0.4591	(	**	)&	0.0495			&	0.0880			&	-0.2335	(	***	)&	0.0273			&	0.4553	(	***	)&	-0.4469	(	***	)&	-0.1794	(	***	)&	-0.1175	(	***	)&	0.5810	(	***	)&	-0.1488	(	***	)\\		
				\bottomrule
		\end{tabular}}
		\begin{minipage}{\linewidth}
			\footnotesize
			\textit{Notes:} This table presents estimated single index coefficients using the full sample period from July 2003 to August 2020, with *** denoting significance at $1\%$ level, ** denoting significance at the $5\%$ level and * denoting significance at the $10\%$ level.
		\end{minipage}
	\end{table}

According to Figure~\ref{figure:ft_theta_q5}, at the lower quantile, e.g., $\tau = 0.05$, three latent factors are identified, characterizing return dynamics for bonds that experience relatively poor performance. Factor 1, which remains relatively stable with moderate disturbances from the 2008 Global Financial Crisis (GFC) until 2012, likely reflects a baseline macroeconomic or broad credit risk component, affecting underperforming bond returns persistently. Factor 2 shows sharp negative movements during 2008–2009, and is likely to have a strong association with systemic risk and liquidity stress, highlighting bonds' sensitivity to spikes in volatility and deteriorating credit conditions during financial turmoil. Factor 3 exhibits large fluctuations during major crisis periods (2008 GFC and the 2020 pandemic) and noticeable spikes around 2006 and 2015. These idiosyncratic shock events likely reflect the Federal Reserve's monetary tightening and early housing market stress in 2006, as well as the 2015 collapse in oil prices, and anticipated interest rate increases in the U.S. This pattern suggests that the factor may be capturing episodic liquidity strains or macro-financial disruptions that fall short of global recessions, such as sector-specific financing squeezes or regional market dislocations. Together, these factors imply that in the left tail of the bond return distribution, bond underperformance is driven by a combination of credit risk, systemic volatility, and liquidity-induced repricing pressures.

Examining the associated index parameters, the impact of characteristics on factor loadings 1 and 3 are relatively balanced, with moderately distributed loadings across most characteristics. In contrast, Factor 2's index estimates highlight significant negative associations with equity volatility and credit spreads. Our model captures interactions between bond spreads and market uncertainty (proxied by equity volatility), a key factor noted in the literature \citep{bell2024glass}. This is achieved through the model’s inherent ability to account for interactions between characteristics.
Lastly, we observe that none of the index parameters for Factor 3 are marginally significant, suggesting it is likely that none of the characteristics are significant in explaining the loading of Factor 3, namely the true loading $\lambda_k(x_{it}'\theta_k)$ for Factor 3 is flat essentially.\footnote{In this case, the model remains correctly specified if the true $\lambda_k(\cdot)$ is constant. However, the single-index structure provides little explanatory power, as it does not vary with characteristics. So it raises concerns about overparameterization, rather than model misspecification.} This likely reflects the nature of Factor 3, which remains largely inactive except during periods of abrupt macroeconomic shocks. In such episodes, the factor appears to influence bond returns broadly and uniformly, irrespective of firm-level characteristics. As a result, its risk exposure cannot be systematically explained by the observable covariates included in the model.

At the median quantile ($\tau = 0.5$), four latent factors are extracted, capturing the return dynamics under more typical market conditions. Factor 1 evolves smoothly over time with moderate reactions during crises, reflecting baseline firm-specific and credit risk exposures. 
Factor 2, which exhibits pronounced random fluctuations around zero, likely captures idiosyncratic or sector-specific shocks unrelated to broader market trends. 
Factors 3 and 4, while also fluctuating around zero, display increased volatility during major stress periods, suggesting sensitivity to market-wide volatility spillovers and spread dynamics under stress.

Regarding the index parameters, several characteristics stand out across the factors in terms of magnitude and significance level: equity volatility, coupon, spread, bond volatility, and VIX beta. Factor 1 loads heavily and positively on equity volatility, whereas Factor 2 shows a strong negative loading on the same characteristic, indicating contrasting responses to firm-specific risk. Equity momentum exhibits significant negative loadings in Factors 1 and 3, but a strong positive loading in Factor 2, suggesting competing dynamics associated with trend-following behavior. Bond volatility and VIX beta appear prominently in Factors 3 and 4, but with opposite signs, pointing to a trade-off between idiosyncratic and market-wide volatility exposure. Coupon, a key bond fundamental, has significant and heterogeneously signed loadings across all four factors, highlighting its complex role in shaping conditional bond returns at the median quantile.

At the upper quantile ($\tau=0.95$), four latent factors are again identified, capturing complex right-tail return dynamics, which can arise from both favorable market conditions and extreme risk-driven repricing episodes.

Factor 1 exhibits modest fluctuations over time, and its large positive loading on coupon suggests that this factor is likely linked to carry returns, where stable income streams, rather than rapid market repricing, are the dominant contributors to high realized returns.
Factor 2 displays large positive spikes during systemic crises (2008 and 2020), and intermediate surges around 2012 and 2016. Its strong positive loading on equity volatility and negative loading on momentum imply that, during these extreme events, bonds associated with riskier, high-volatility firms experience positive tail returns, while bonds linked to strong past momentum underperform — emphasizing that high returns at the upper tail do not always arise from economic expansions but often reflect complex disorderly market corrections. 
Factor 3 follows a slight upward trend until the 2008 crisis, experiences a sharp dip, and resumes its upward trajectory after 2010, suggestive of a long-term macro-trend factor sensitive to regime shifts. It loads positively on VIX beta and negatively on bond volatility and coupon, capturing systemic volatility exposure with reduced reliance on steady carry. Factor 4 shows a sharp drop around 2010 and a gradual downward trend until 2018, loading negatively on coupon and positively, though moderately, on bond age, bond volatility, duration, face value, and equity market capitalization, suggesting a growth-sensitive dynamic tilt away from carry-dominant bonds. 

Overall, the upper tail factors underscore that outperforming bond returns may emerge from diverse forces, including both carry-driven mechanisms and volatility-driven systemic repricing, cautioning against overly simplistic interpretations of upper tail behavior as purely reflective of a booming market.

Across quantiles, the estimated latent factors capture distinct aspects of bond return dynamics: systematic credit risk and liquidity shocks at the lower tail, normal firm-specific and volatility risks at the median, and a mix of carry-driven returns and systemic repricing effects at the upper tail. The associated index parameters provide valuable structural information about how characteristics shape latent factor exposures distinctively across factors and quantiles, although the nonlinearity of the loading functions necessitates caution when interpreting their marginal effects. Together, the results reveal complex, state-dependent interactions between bond characteristics and latent risk factors across the full distribution of returns.


	\section{Conclusion}\label{sec:conclusion}

In this paper, we propose a characteristics-augmented quantile factor (QCF) model that links latent factor loadings to observable characteristics via a single-index structure. We develop a three-step estimation procedure and establish its theoretical properties, including consistency and asymptotic normality under mild conditions. Simulation and empirical results demonstrate the model’s effectiveness.

Applying the QCF model to U.S. corporate bond returns, we show that it outperforms benchmark models, including quantile factor models (QFM) and quantile regressions based on the Fama-French 5 factors (QR-FF5), especially in the tails of the return distribution. By capturing nonlinear, covariate-dependent factor structures, the QCF model improves both in-sample fit and out-of-sample predictive accuracy. While QFM performs comparably near the median, QCF offers greater flexibility and explanatory power under extreme market conditions. The estimated latent factors admit clear economic interpretations across quantiles, and the index structure provides structural insights into how bond characteristics shape factor exposures in a state-dependent manner.

This paper focuses on the case where factor loadings are fully explained by observed characteristics. An important direction for future research is to allow for partially characteristic-driven loadings of the form:
\begin{align*}
    Q_{y_{it}}(\tau | \mathcal{F}_{t-1}) = \sum_{k=1}^r f_{t(k)}\lambda_{it(k)}, \quad \lambda_{it(k)} = \lambda_{i(k)} + \lambda_k(x_{it}'\theta_k).
\end{align*}
where $\lambda_{i(k)}$ is unobserved and independent of $x_{it}$. Under the mean framework, \citet{fan2016projected} considers a similar extension, allowing factor loadings to depend on unobserved heterogeneity. Extending this to the quantile setting is nontrivial due to the lack of summability in quantile restrictions. Developing such models would be a meaningful and technically challenging direction for future work. 
	\newpage
	\bibliographystyle{agsm}
	\bibliography{mybibFile}

	\newpage
\appendix

    \section*{Appendix}
\setcounter{equation}{0} 
\setcounter{lemma}{0}
\setcounter{page}{1}
\setcounter{proposition}{0} 
\setcounter{section}{0}
\setcounter{lemma}{0}
\setcounter{theorem}{0}

\renewcommand{\thepage}{A-\arabic{page}}
\renewcommand{\theequation}{A.\arabic{equation}}
\renewcommand{\thesection}{A.\arabic{section}}
\renewcommand{\thefigure}{A.\arabic{figure}}
\renewcommand{\thetable}{A.\arabic{table}}
\renewcommand{\theassumption}{A.\arabic{assumption}} 
\renewcommand{\theproposition}{A.\arabic{proposition}} 
\renewcommand{\thelemma}{A.\arabic{lemma}}
\renewcommand{\thetheorem}{A.\arabic{theorem}}
\renewcommand{\thecorollary}{A.\arabic{corollary}}

In this appendix, Section~\ref{appendix:identification} verifies the factor identification condition in Assumption~\ref{ass.identification}. 
Section~\ref{appendix:proofA} establishes the proofs of the theorems in the main text. Section~\ref{appendix:lemma} states and proves the preliminary lemmas required in Section~\ref{appendix:proofA}.

\ifarxiv
\section{Understanding the Identification Conditions}\label{appendix:identification}
\section{Proof of the Main Results}\label{appendix:proofA}
\section{Preliminary Lemmas}\label{appendix:lemma}
For full proofs and preliminary lemmas, see the full version.
\else
\input{Appendix2_3}
\fi

\ifarxiv
\newcounter{dummyprop}
\renewcommand{\thedummyprop}{A.1}
\refstepcounter{dummyprop}
\label{t1:bahadur}
\fi

\end{document}